%% file: paper.tex
\documentclass[sigconf, screen, nonacm]{acmart}

\AtBeginDocument{%
  }

\usepackage{tikz}
\usepackage{xspace}
\usepackage{siunitx}
\usepackage{amsmath}
\usepackage{subfig}
\usepackage{multicol}
\usepackage{soul}

\usepackage[]{hyperref}

\input{macro}
\graphicspath{{figs/}}

\begin{document}

\title{\proj: Real-Time Mobile Neural Rendering by Exploiting Computational Redundancy}

\input{author}


\input{abstract}



\maketitle 

\input{intro}
\input{motiv}
\input{algo}

\input{cache}
\input{arch}
\input{exp}
\input{eval}
\input{related}
\input{discussion}
\input{conclusion}

\bibliographystyle{plain}
\bibliography{references}

\end{document}

%% file: macro.tex
\newcommand*\circled[2]{\tikz[baseline=(char.base)]{
            \node[shape=circle,fill=black,inner sep=1pt] (char) {\textcolor{#1}{{\footnotesize #2}}};}}

\ifx\figurename\undefined \def\figurename{Figure}\fi
\renewcommand{\figurename}{Fig.}
\renewcommand{\paragraph}[1]{\textbf{#1} }

\newcommand{\Sect}[1]{Sec.~\ref{#1}}
\newcommand{\Fig}[1]{Fig.~\ref{#1}}

\newcommand{\Eqn}[1]{Eqn.~\ref{#1}}

\def\cS{{\mathcal{S}}}
\def\cRC{{\mathcal{RC}}}

\newcommand{\proj}{\textsc{Lumina}\xspace}
\newcommand{\sys}{\textsc{LuminSys}\xspace}
\newcommand{\core}{\textsc{LuminCore}\xspace}
\newcommand{\cache}{\textsc{LuminCache}\xspace}
\newcommand{\algo}{$\cS^2$\xspace}
\newcommand{\rc}{$\cRC$\xspace}
\newcommand{\mode}[1]{\underline{\textsc{#1}}\xspace}

\newcommand{\RNum}[1]{\uppercase\expandafter{\romannumeral #1\relax}}


%% file: author.tex
\author{Yu Feng}
\orcid{0000-0002-2192-5737}
\affiliation{%
  \institution{Shanghai Jiao Tong University, Shanghai Qi Zhi Institute}
  \city{Shanghai}
  \country{China}
}
\email{y-feng@sjtu.edu.cn}

\author{Weikai Lin}
\orcid{0000-0003-3537-4857}
\affiliation{%
  \institution{University of Rochester}
  \city{Rochester}
  \state{NY}
  \country{USA}
}
\email{wlin33@ur.rochester.edu}

\author{Yuge Cheng}
\orcid{0009-0001-5587-0463}
\affiliation{%
  \institution{Shanghai Jiao Tong University}
  \city{Shanghai}
  \country{China}
}
\email{chengyuge@sjtu.edu.cn}

\author{Zihan Liu}
\orcid{0000-0002-0874-0682}
\affiliation{%
  \institution{Shanghai Jiao Tong University, Shanghai Qi Zhi Institute}
  \city{Shanghai}
  \country{China}
}
\email{altair.liu@sjtu.edu.cn}

\author{Jingwen Leng}
\authornote{Corresponding Authors.}
\orcid{0000-0002-5660-5493}
\affiliation{%
  \institution{Shanghai Jiao Tong University, Shanghai Qi Zhi Institute}
  \city{Shanghai}
  \country{China}
}
\email{leng-jw@sjtu.edu.cn}

\author{Minyi Guo}
\authornotemark[1]
\orcid{0000-0003-0034-2302}
\affiliation{%
  \institution{Shanghai Jiao Tong University, Shanghai Qi Zhi Institute}
  \city{Shanghai}
  \country{China}
}
\email{guo-my@sjtu.edu.cn}

\author{Chen Chen}
\orcid{}
\affiliation{%
  \institution{Shanghai Jiao Tong University}
  \city{Shanghai}
  \country{China}
}
\email{chen-chen@sjtu.edu.cn}

\author{Shixuan Sun}
\orcid{}
\affiliation{%
  \institution{Shanghai Jiao Tong University}
  \city{Shanghai}
  \country{China}
}
\email{sunshixuan@sjtu.edu.cn}

\author{Yuhao Zhu}
\orcid{0000-0002-2802-0578}
\affiliation{%
  \institution{University of Rochester}
  \city{Rochester}
  \state{NY}
  \country{USA}
}
\email{yzhu@rochester.edu}

%% file: abstract.tex
\begin{abstract}

3D Gaussian Splatting (3DGS) has vastly advanced the pace of neural rendering, but it remains computationally demanding on today's mobile SoCs.
To address this challenge, we propose \proj, a hardware-algorithm co-designed system, which integrates two principal optimizations: a novel algorithm, \algo, and a radiance caching mechanism, \rc, to improve the efficiency of neural rendering. 
\algo algorithm exploits temporal coherence in rendering to reduce the computational overhead, while \rc leverages the color integration process of 3DGS to decrease the frequency of intensive rasterization computations.
Coupled with these techniques, we propose an accelerator architecture, \core, to further accelerate cache lookup and address the fundamental inefficiencies in Rasterization. 
We show that \proj achieves $4.5\times$ speedup and 5.3$\times$ energy reduction against a mobile Volta GPU, with a marginal quality loss ($<$ 0.2~dB peak signal-to-noise ratio reduction) across synthetic and real-world datasets.

\end{abstract}

%% file: intro.tex
\section{Introduction}
\label{sec:intro}



Neural Radiance Fields (NeRF)~\cite{mildenhall2021nerf} has transformed the landscapes of Virtual/Augmented Reality (VR/AR)~\cite{rojas2023re, chen2023mobilenerf, hu2022efficientnerf, hedman2021baking}, large-scale landscape modeling~\cite{xiangli2022bungeenerf, barron2021mip, lindell2022bacon, tancik2022block}, virtual avatar~\cite{chen2022geometry, jiang2022neuman, weng2022humannerf}, novel view synthesis~\cite{ye2023intrinsicnerf, zhang2022ray, liang2023envidr}, and beyond~\cite{gao2022nerf}.
Although NeRF yields impressive results, its intensive computation is always a key bottleneck to achieving real-time and high-resolution rendering~\cite{wu2024recent, chen2024survey}.

To address this challenge, 3D Gaussian Splatting (3DGS) has been proposed as a fast alternative to the NeRF pipeline~\cite{kerbl20233d}. 
Unlike NeRF, which requires dense sampling along each ray, 3DGS projects precomputed Gaussian points onto the rendering screen, simplifying the color integration process and light transport modeling in 3D spaces. 
Despite numerous efforts~\cite{kerbl20233d, wu2024recent, chen2024survey, fang2024mini, lee2023compact, fan2023lightgaussian}, 3DGS still falls short of the real-time requirement, e.g., 90 frames per second (FPS)~\cite{questprospec, visionprospec, wang2023effect}, in AR/VR applications. 
For instance, on a mobile Volta GPU~\cite{xaviersoc}, 3DGS merely achieves 5 - 21 FPS on real-world scenes~\cite{barron2022mipnerf360, Knapitsch2017, hedman2018deep}, far from the real-time target.

We aim to enhance the efficiency of 3DGS rendering in AR/VR applications. 
Our characterizations show that Sorting and Rasterization dominate the 3DGS rendering time, accounting for over 90\% of the total execution (\Sect{sec:mot:perf}).
To address the two main bottlenecks in 3DGS, we introduce \proj, a hardware-algorithm co-designed system in this paper.

\begin{figure*}[t]
\centering
\includegraphics[width=\textwidth]{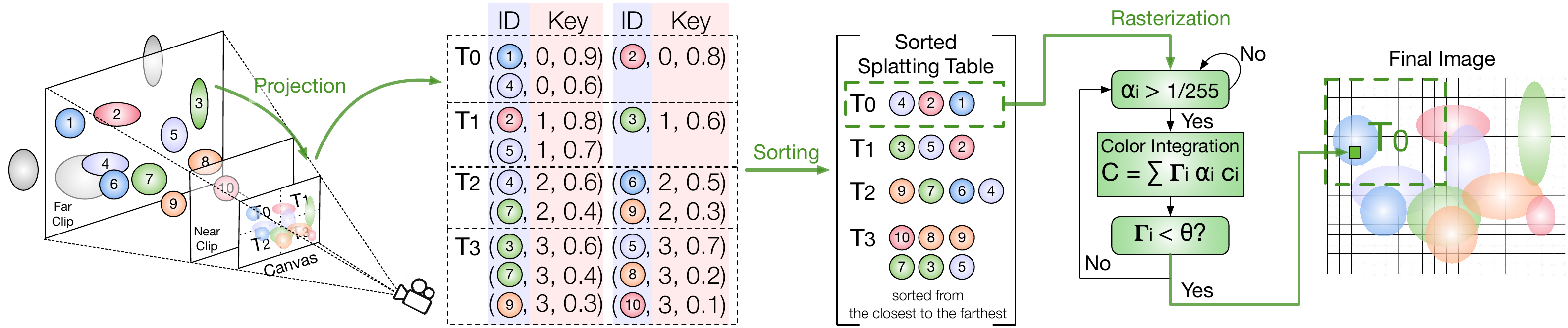}
\caption{The computation flow of today’s 3DGS rendering pipeline is highlighted in green. Gaussian points are first projected onto the rendering screen to determine their intersection with image tiles (e.g., $T_0$, $T_1$). Next, each tile sorts the intersected points based on their depth values, from the nearest to the farthest. Finally, within each tile, Gaussian points are splatted onto the screen in the sorted order to render the final image governed by the color integration equation (\Eqn{eqn:nerf}). Note that, only those Gaussians with significant transparency $\alpha_i$ ($\alpha_i > \frac{1}{255}$) will be integrated in the final pixel.}
\label{fig:algo_flow}
\end{figure*}

\paragraph{Algorithms.} We first propose a novel \textit{Sorting-Shared} (\algo) algorithm to address one of the bottlenecks, Sorting.
\algo algorithm extends 3DGS by leveraging temporal redundancy in rendering (\Sect{sec:algo:ss}), sharing sorting results across multiple frames.
In addition, \algo preemptively executes Sorting, effectively hiding Sorting execution time and mitigating frame stuttering.
By doing so, \proj reduces the Sorting computations without losing quality.

While \algo algorithm effectively hides Sorting latency in 3DGS, it does leave the other bottleneck, Rasterization, which is compute-intensive due to its color integration process. 
To accelerate Rasterization, we propose \textit{Radiance Caching} (\rc), a lightweight caching mechanism that leverages the previous rendering results to approximate the current pixel values with a negligible compute overhead.

\rc leverages a key insight: \textit{two rays intersecting the same sequence of Gaussian points would likely yield the same pixel values} (\Sect{sec:algo:rc}).
Specifically, we propose to cache a short record of ``significant'' ray-Gaussian intersections from previous renderings. 
Instead of executing the whole color integration, this short record allows us to perform only the initial segment of the color integration and determine the pixel values by matching with the cached record.

\paragraph{Fine-Tuning.} However, directly applying \rc to 3DGS occasionally introduces rendering artifacts.
To address this, we propose end-to-end fine-tuning for our new pipeline to enhance rendering quality while maintaining the caching efficiency.
Overall, our result shows that \rc avoids 55\% computation, on average, in color integration with minimal impact on quality.

\paragraph{Architectural Support.} Coupled with two optimizations, we further propose a customized architecture, \core (\Sect{sec:arch}), to address the inherent computation inefficiencies in Rasterization. 
Our characterization shows that the color integration in Rasterization is inherently sparse, leading to low GPU utilization due to warp divergence.
Radiance-cached Rasterization further exacerbates this divergence, transforming a full pixel rendering into a sparser one.
Thus, we design dedicated Neural Rendering Units (NRUs) in \core to tackle the sparse color integration in Rasterization. 
Along with NRUs, we co-design a specialized hardware cache, \cache, to speed up the cache lookup in \rc. 

\paragraph{Result.} By integrating \proj into an off-the-shelf mobile SoC with negligible hardware overhead (0.4\%), we show that \proj achieves $4.5\times$ speedup and 5.3$\times$ energy efficiency against a mobile Volta GPU, with a marginal quality loss. 

Our contributions are summarized as follows:
\begin{itemize}
    \item We introduce a plug-and-play \algo algorithm that leverages the temporal coherence in 3DGS to completely hide the Sorting overhead during real-time rendering.
    \item We propose a novel \rc mechanism that caches ray-Gaussian intersections and reduces the computation of color integration in Rasterization by 55\%.  
    \item We propose a \core architecture, co-designed to address the GPU warp divergence in Rasterization.
    \item We demonstrate that \proj can achieve $4.5\times$ speedup and 5.3$\times$ energy reduction against a mobile Volta GPU, with a marginal loss on visual quality.
\end{itemize}

%% file: motiv.tex
\section{Motivation}
\label{sec:mot}

We begin by introducing 3DGS fundamentals in \Sect{sec:mot:pipe}.
Then, we identify the main inefficiencies of the 3DGS pipeline in \Sect{sec:mot:perf}.

\subsection{3DGS Pipeline}
\label{sec:mot:pipe}

\paragraph{Why 3DGS?} Recent 3DGS~\cite{kerbl20233d} revolutionizes NeRF rendering by drastically accelerating the conventional NeRF rendering process.
Conventional NeRF rendering is known for its computational intensity due to the extensive ray samplings. 
Each ray sample is required to perform a MLP operation.

In contrast, 3DGS proposes a reverse operation called ``splatting'', where Gaussian points (or ``Gaussians'' for short) are directly projected onto the rendering screen. 
This method sidesteps the compute-heavy task of ray-object intersections by reversing the workflow: rather than rays seeking Gaussians, Gaussians are directly mapped onto the screen.

To date, all 3DGS variants adhere to a \textit{uniform} rendering process~\cite{kerbl20233d, wu2024recent, chen2024survey, fang2024mini, lee2023compact, fan2023lightgaussian}.
The primary variation across different 3DGS algorithms is the training methodology, \textit{not} the rendering process. 
Thus, this paper focuses on the original 3DGS rendering process~\cite{kerbl20233d} without loss of generality.

As \Fig{fig:algo_flow} shows, 3DGS executes rendering tile-by-tile through three steps: \textit{Projection}, \textit{Sorting}, and \textit{Rasterization}.

\paragraph{Projection.}
Given a camera pose, Projection serves two main purposes: first, it filters out Gaussians that fall outside the view frustum (e.g. grey ellipsoids), retaining only those Gaussians between the near- and far-clip planes (e.g. colored ellipsoids) as shown in \Fig{fig:algo_flow};
second, it projects each Gaussian, with a defined cutoff radius, onto the screen to determine its intersecting tiles. 



\paragraph{Sorting.} Once each tile collects its intersected Gaussian IDs, Sorting then determines the rendering order of those Gaussians, ensuring that all points are rendered according to their depth, from the closest to the furthest (relative to the camera pose), as shown in Sorted Splatting Table. 

\paragraph{Rasterization.} Once all Gaussians are sorted, Rasterization renders these Gaussians tile-by-tile. 
Every pixel within a tile would iterate the same set of Gaussians, calculating the transparency and integrating the color of those Gaussians to its pixel in the sorted order. 
For example, every pixel in tile $T_0$ integrates Gaussians in \circled{white}{4} $\rightarrow$ \circled{white}{2} $\rightarrow$ \circled{white}{1} order. \Eqn{eqn:nerf} governs the color integration of pixel $\textbf{p}$:
\begin{align}
\label{eqn:nerf}
   C(\textbf{p}) & = \sum_{i=1}^{N} \Gamma_i \alpha_i \textbf{c}_{i},\ where \ \Gamma_i = \prod^{i-1}_{j=1} (1-\alpha_j)
\end{align}
where $\Gamma_i$ denotes the accumulative transmittance of pixel $\textbf{p}$ from the first Gaussian $1$st to the $i-1$th. $\alpha_i$ and $\textbf{c}_i$ stand for the transparency and the color at the $i$th Gaussian, respectively.

Note that, if the Gaussian's $\alpha_i$ value is smaller than $\frac{1}{255}$, this Gaussian will be skipped in the color integration process to avoid numerical instabilities, as shown in \Fig{fig:algo_flow}. 
The color integration terminates once the accumulative transmittance $\Gamma_i$ falls below a predefined threshold, $\theta$.
\subsection{Performance Characterization}
\label{sec:mot:perf}

\begin{figure}[t]
\centering
\subfloat[Model size.]{
	\label{fig:model_size}	
        \includegraphics[width=0.47\columnwidth]{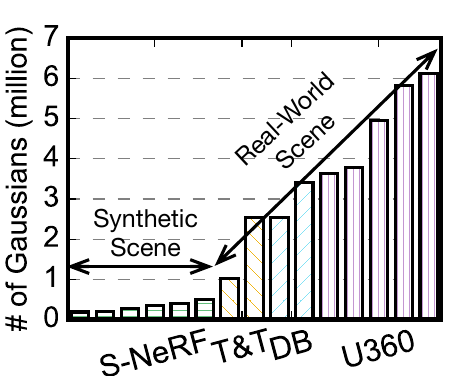} 
}
\subfloat[Frame-per-second.]{
	\label{fig:latency}
	\includegraphics[width=0.48\columnwidth]{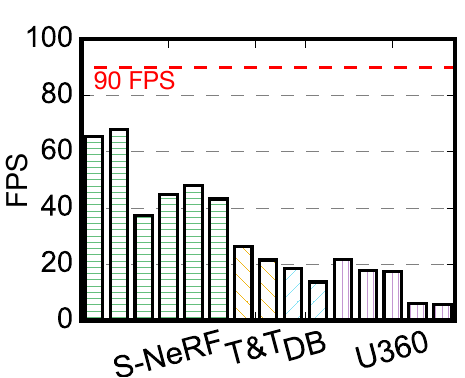} 
 } 
\caption{Model size and runtime performance trends with scene complexity. 
Rendering real-world scenes requires longer latencies, driven by their increased scene complexity. 
Performance measurements conducted across four key datasets: S-NeRF~\cite{mildenhall2021nerf}, T\&T~\cite{Knapitsch2017}, DB~\cite{hedman2018deep}, and U360~\cite{barron2022mipnerf360}.}
\label{fig:correlation}
\end{figure}

\paragraph{Model Size.} To achieve photorealism, 3DGSs require dense Gaussians to reconstruct the scene. 
\Fig{fig:model_size} shows the correlation between scene complexity and model size. 
Moving from small-scale synthetic to real-world scenes, there is a significant increase in model size. 
Synthetic scenes in the S-NeRF dataset~\cite{mildenhall2021nerf} contain relatively few Gaussians (less than 1 million). 
However, real-world datasets~\cite{Knapitsch2017, hedman2018deep, barron2022mipnerf360} increase the number of Gaussians over 10$\times$, with the U360 dataset reaching over 6 million Gaussians. 

\paragraph{Performance.}
For real-world scenes, the growth in model size poses computational challenges compared to their synthetic counterparts. 
As more detailed rendering is demanded in today's applications, increasingly large-scale models~\cite{kerbl2024hierarchical, liu2024citygaussian} are necessitated for the intricate geometries and light interactions in the real world.
\Fig{fig:latency} shows the rendering performance across datasets on a mobile Volta GPU on Nvidia Xavier SoC~\footnote{The performance of the Volta GPU (2.8 TFLOPS) is comparable to that of the GPU (3.5 TFLOPS) in the Snapdragon XR2~\cite{snapdragonxr2}, released in September 2023 for VR platforms~\cite{xr2release}.}~\cite{xaviersoc}. 
As the rendering scenes transition from synthetic to real-world scenarios, the performance drops from 66 to 5 FPS, far below the real-time requirements of 90-100 FPS for AR/VR platforms~\cite{questprospec, visionprospec}. 



\begin{figure}[t]
\centering
\begin{minipage}[t]{0.48\columnwidth}
  \centering
  \includegraphics[width=\columnwidth]{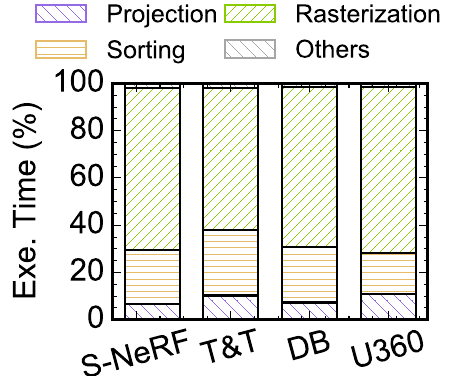}
  \caption{Normalized execution breakdown across scenes. Sorting and Rasterization dominate the execution.}
  \label{fig:exec_time}
\end{minipage}
\hspace{2pt}
\begin{minipage}[t]{0.48\columnwidth}
  \centering
  \includegraphics[width=\columnwidth]{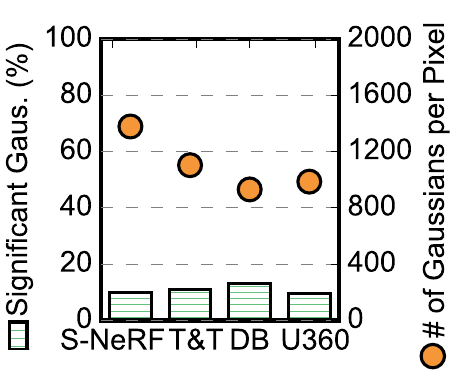}
  \caption{The percentage of significant Gaussians per pixel and the average number of iterated Gaussians per pixel.}
  \label{fig:data_util}
\end{minipage}
\end{figure}

\paragraph{Execution Breakdown.} To understand the main bottlenecks in 3DGS, \Fig{fig:exec_time} presents the normalized execution breakdown across four datasets.
Overall, Sorting and Rasterization dominate the execution time, accounting for 23\% and 67\% on average, respectively.
One thing worth mentioning, as the model size increases, no clear trend shows significant changes in the execution distribution.
This means that the optimization focus will not shift as the scene scales.

\paragraph{Sparse Color Integration.} 
With Rasterization dominating the overall execution time, we further pinpoint its performance bottleneck.
By default, pixels within a tile are designed to iterate the same set of Gaussians, as shown in \Fig{fig:algo_flow}. 
However, Gaussians contribute to the final pixel in \Eqn{eqn:nerf} only if their transparencies, $\alpha$, exceed $\frac{1}{255}$.
We refer to these Gaussians as \textit{significant Gaussians}.
\Fig{fig:data_util} characterizes the percentage of significant Gaussians versus the total Gaussians iterated per pixel across four datasets in Rasterization.

In \Fig{fig:data_util}, the left y-axis shows the percentage of significant Gaussians while the right y-axis shows the average number of total Gaussians per pixel. 
Despite each pixel iterating over a thousand of Gaussians, the percentage of significant Gaussians remains low, averaging only 10.3\%
with a standard deviation of 2.1\%. 
This shows that each pixel is generated by only a small subset of Gaussians.





\begin{figure}[t]
    \centering
    \includegraphics[width=\columnwidth]{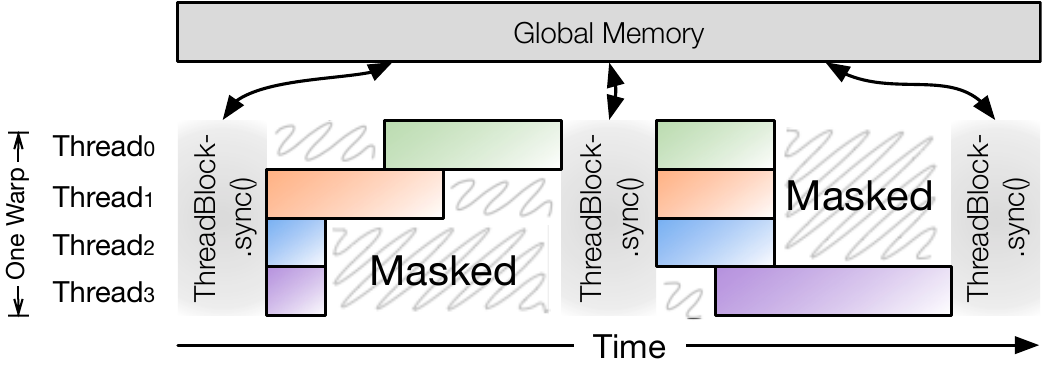}
    \caption{Example of a warp execution during Rasterization, assuming 4 threads per warp. 
    Colored blocks show that threads are not masked and perform meaningful computations. 
    Between two computation periods, all threads load data from global memory and synchronize.
    Sparse color integration shown in \Fig{fig:data_util} leads to severe warp divergence.
    }
    \label{fig:thread_idle_exp}
\end{figure}

\paragraph{Warp Divergence.}
Due to the sparsity in the color integration (\Fig{fig:data_util}), different pixels might integrate different subsets of Gaussians.
However, in typical GPU implementations, Rasterization is parallelized across pixels, assigning one thread to one pixel. 
The workload discrepancy across pixels would result in inefficient GPU utilization due to warp divergence in \Fig{fig:thread_idle_exp}.
Since threads are grouped into warps in modern GPU, all threads in a warp execute together in a \textit{Single Instruction Multiple Thread} (SIMT) fashion.

The example in \Fig{fig:thread_idle_exp} shows a simplified GPU execution model with four threads within a warp.
Here, one color represents one thread activity during rendering.
The color integrations are interleaved with data synchronization, i.e., fetching Gaussians from global memory to shared memory. 
Since each thread requires different Gaussians, the GPU will mask threads that do not need to integrate specific Gaussians at that time. 
Our result shows that threads remain masked over 69\% of the time across scenes with a standard deviation of 10\%, showing low GPU utilization. 
This inefficiency underscores that GPUs are \textit{ill-suited} for Rasterization.


Although prior studies propose various techniques to address GPU warp divergence~\cite{meng2010dynamic, fung2007dynamic, narasiman2011improving}, these methods are not suited for Rasterization in 3DGS. 
Techniques such as dynamic warp formation~\cite{fung2007dynamic} and dynamic warp subdivision~\cite{meng2010dynamic} rely on dynamic scheduling to mitigate warp divergence but inherently require synchronization, introducing additional runtime scheduling overhead.
To overcome these limitations, we propose customized hardware for Rasterization (in \Sect{sec:arch}) that eliminates synchronization overhead while maintaining high utilization of hardware resources.

%% file: algo.tex
\section{\sys}
\label{sec:algo}

This section presents our system, \sys. We first introduce two plug-and-play optimizations, \textit{Sorting Sharing} (\Sect{sec:algo:ss}) and \textit{Radiance Caching} (\Sect{sec:algo:rc}), to address two main bottlenecks, Sorting and Rasterization, respectively. 
We then explain how \sys integrates the optimizations to guarantee a smooth rendering (\Sect{sec:algo:sys}).

\subsection{Sorting Sharing, \algo}
\label{sec:algo:ss}

\begin{figure}[t]
\centering
\includegraphics[width=\columnwidth]{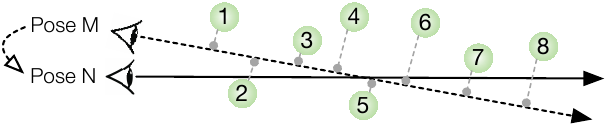}
\caption{The intuition of \algo algorithm. The rendering orders, a.k.a, the depth orders, of two spatially closed camera poses, $M$ and $N$, are the same and can be reused. 
}
\label{fig:ss_intuition}
\end{figure}

\paragraph{Intuition.} The high-level idea of our \algo algorithm is to reuse the previous sorting result and bypass Sorting in 3DGS.
The rationale here is that the sorting results tend to remain the same across consecutive poses.
\Fig{fig:ss_intuition} gives an example.
As the camera moves from pose~$M$ to pose~$N$, the depth order of these Gaussians remains unchanged. 
Numbers in \Fig{fig:ss_intuition} show the depth order.
The key observation here is that the sequence in which these Gaussians get rendered does not require frequent re-computation. 
This allows us to skip Sorting in adjacent frames.

Even if these Gaussians' rendering order does change locally during rendering, the locally incorrect order is unlikely to impact the overall rendering results. 
This stability comes from the sparsity of color integration characterized in \Sect{sec:mot:perf}.
Significant Gaussians (whose $\alpha > \frac{1}{255}$) of each pixel are likely separated apart after sorting, and their relative order will not change too much from one camera pose to another.
Our experiment shows that only 0.2\% of these Gaussian orders are changed.

 


\begin{figure}[t]
\centering
\includegraphics[width=\columnwidth]{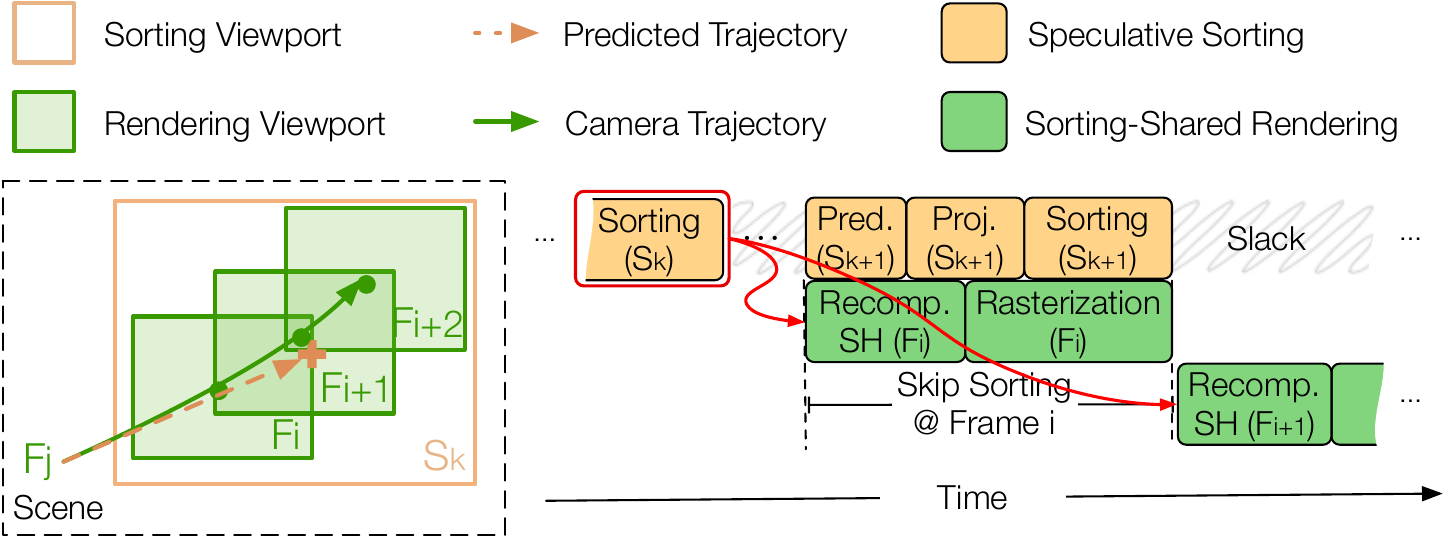}
\caption{Our \algo algorithm consists of two concurrent paths: \textit{speculative sorting} and \textit{sorting-shared rendering}. Speculative sorting predicts the future pose (e.g., $S_k$) and preemptively sorts Gaussians at that predicted pose.
Note that, the sorting viewport at $S_k$ requires to be large enough to accommodate all rendering ports in the sharing window, i.e., $F_i$, $F_{i+1}$, and $F_{i+2}$.
Sorting-shared rendering later bypasses Sorting and performs Rasterization directly.
}
\label{fig:ss_algo}
\end{figure}

\paragraph{Algorithm.}
One potential issue with a naive skip-sorting strategy is that we still need to perform Sorting in the middle of our rendering process.
This inevitably introduces frame stuttering due to additional computation from Sorting.
To hide the Sorting latency, we propose our \algo algorithm to keep a more consistent delay across frames.
The key idea of \algo is to predict the pose trajectory and pre-compute the sorting results in advance.

\Fig{fig:ss_algo} illustrates the workflow of our \algo algorithm when rendering a scene across different camera poses over time. 
Our \algo algorithm consists of two concurrent execution paths: \textit{speculative sorting} and \textit{sorting-shared rendering}.

\underline{\textit{Speculative Sorting.}} 
Speculative sorting is to predict the future camera pose and pre-sort Gaussians at this predicted pose before actual rendering.
\Fig{fig:ss_algo} highlights the speculative sorting in yellow.
The red arrows highlight the data dependency between Sorting and Rasterization.
At pose $F_j$, the \algo algorithm first predicts a future pose $S_k$ using the current position and velocity at $F_j$.
The velocity at $F_j$ is approximated from the last two camera poses, $F_{j-1}$ and $F_j$:
\begin{align}
v_j = \frac{F_j -  F_{j-1}}{\Delta t},
\end{align}
where $\Delta t$ is the interval between two consecutive frames. The future pose $S_k$ is then predicted using:
\begin{align}
	S_k = T_k + v \times t_r,~~~t_r = \frac{N}{2}\Delta t,
\end{align}
where $N$ is the number of rendered frames that share the same sorting result.
We define $N$ as \textit{sharing window}.
Using $\frac{N}{2}$ allows the predicted pose to be approximately at the center of its related frames, increasing the overlap of the sorted results with its rendered frames.
Once the future pose of $S_k$ is predicted, the \algo algorithm performs the Projection and Sorting steps and saves the sorting result for later use.
Note that, the trajectory prediction is similar to that used in Cicero\mbox{~\cite{feng2024cicero}}. 
We do \textit{not} claim this as our contribution.


\underline{\textit{Sorting-Shared Rendering.}} \Fig{fig:ss_algo} colors the sorting-shared rendering in green. 
During rendering, as the camera moves to a new pose $F_i$, our \algo algorithm skips the repetitive Projection and Sorting steps by reusing the previous sorting result at the predicted pose $S_k$.
One caution here is that, before Rasterization, each Gaussian color needs to be recalculated using pretrained Spherical Harmonic coefficients to preserve view-dependent rendering accuracy.

The subsequent frames (e.g., $F_{i+1}$ and $F_{i+2}$ in \Fig{fig:ss_algo}) within the same sharing window reuse the same sorting result and only perform sorting-shared rendering rather than the entire 3DGS pipeline.
We empirically find that applying a fixed-size sharing window yields a good rendering quality, while prior works~\cite{zhu2018euphrates, han2020megatrack} have exploited various size-adaptive strategies, which are not the \textit{main} focus of this work.
Our evaluation discusses the sensitivity of \algo to the window size in \Sect{sec:eval:sens}.


\begin{figure}[t]
\centering
\includegraphics[width=\columnwidth]{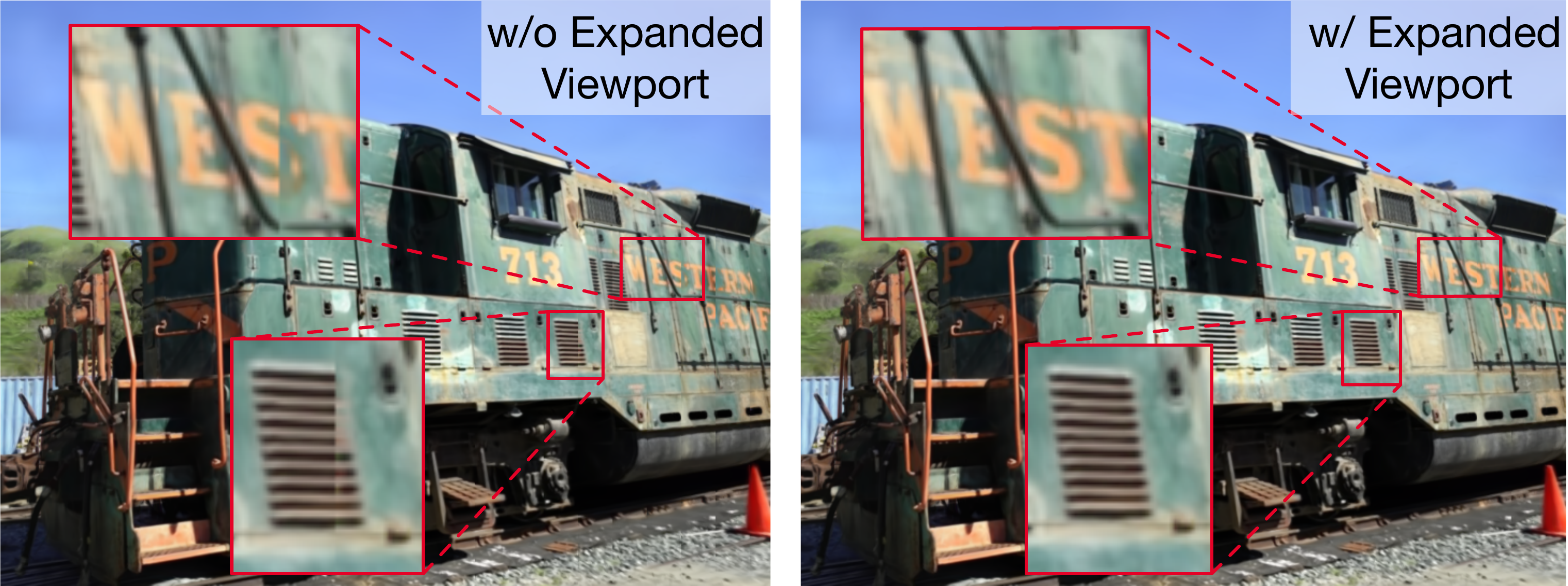}
\caption{Comparison of the rendering results with and without the expanded viewport. The one without the expanded viewport has noticeable artifacts at the tile edges (across `S').}
\label{fig:sort_artifact}
\vspace{-5pt}
\end{figure}

\paragraph{Expanded Viewport.} 
During rendering, each pose is associated with a viewport that includes all the points that can potentially contribute to the pixels under that pose, as shown in \Fig{fig:ss_algo}.
This means that the sorting viewport, $S_k$, needs to cover all the rendering viewports in its sharing window.
If the sorting viewport is too small, Gaussians outside the sorting viewport can introduce rendering artifacts.
As illustrated in the left part of \Fig{fig:sort_artifact}, the rendered image on the left has artifacts across the letter `S'.

To mitigate this issue, we expand the sorting viewport at the speculative sorting stage as outlined by the yellow box $S_k$ in \Fig{fig:ss_algo}.
The viewport at $S_k$ is required to accommodate all viewports in the rendered frames, i.e., $F_i$, $F_{i+1}$, and $F_{i+2}$, in the sharing window.
In practice, since 3DGS operates on a tile-by-tile basis, our \algo algorithm also expands the viewport at a tile granularity.
\Sect{sec:eval:sens} provides a quantitative analysis of how the expanded viewport impacts overall performance.


%% file: cache.tex
\subsection{Radiance Caching, \rc}
\label{sec:algo:rc}

\begin{figure}[t]
\centering
\includegraphics[width=\columnwidth]{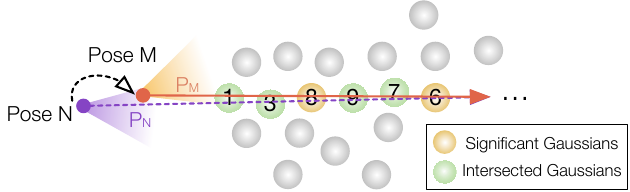}
\caption{The intuition behind radiance caching. The first six intersected Gaussians by pixel $P_N$ and pixel $P_M$ are the same, thus, their ray directions are similar.
The previously rendered pixel $P_N$ can be used to approximate pixel $P_M$.
}
\label{fig:rc_intuition}
\end{figure}

While \algo can hide the execution time of Sorting, Rasterization itself is still the most dominant step in 3DGS.
Here, we propose our technique, \textit{radiance caching}, to reduce the computation in Rasterization. 
Although our technique shares the name with prior techniques in ray tracing~\cite{krivanek2005radiance, vardis2014real, muller2021real}, our technique fundamentally differs from conventional radiance cache in ray tracing as discussed in \Sect{sec:related}.



\paragraph{Intuition.} We begin by explaining the intuition behind our technique. 
One fundamental property of 3DGS is that two rays will yield approximately the same pixel values if they meet two conditions: 1) they share the same direction, and 2) they intersect the same sequence of Gaussians.


The challenge lies in how to identify such rays and thus avoid redundant computations.
Recall in geometry that any two distinct points on a straight line completely determine that line.
Therefore, if two rays intersect with two distinct and sufficiently small Gaussians, these two rays satisfy the two requirements above and, thus, the pixel value of one ray can effectively approximate that of the other. 
The more Gaussians are intersected and shared by two rays, the more confident we are to approximate one using the other.

\Fig{fig:rc_intuition} illustrates our idea. 
Here, poses $M$ and $N$ are close to each other.
In \Fig{fig:rc_intuition}, for pixel $P_N$ at pose $N$ and pixel $P_M$ at pose $M$, their first six intersected Gaussians are identical.
Although the remaining thousands of Gaussian intersections for each pixel have not yet been computed, the similarity in their initial intersected Gaussians allows us to approximate one pixel value using the other.
By saving the pixel value of $P_N$ from pose $N$, the Rasterization step of $P_M$ can safely terminate at the sixth Gaussian and use $P_N$’s pixel value to approximate that of $P_M$.

\begin{figure}[t]
\centering
\includegraphics[width=\columnwidth]{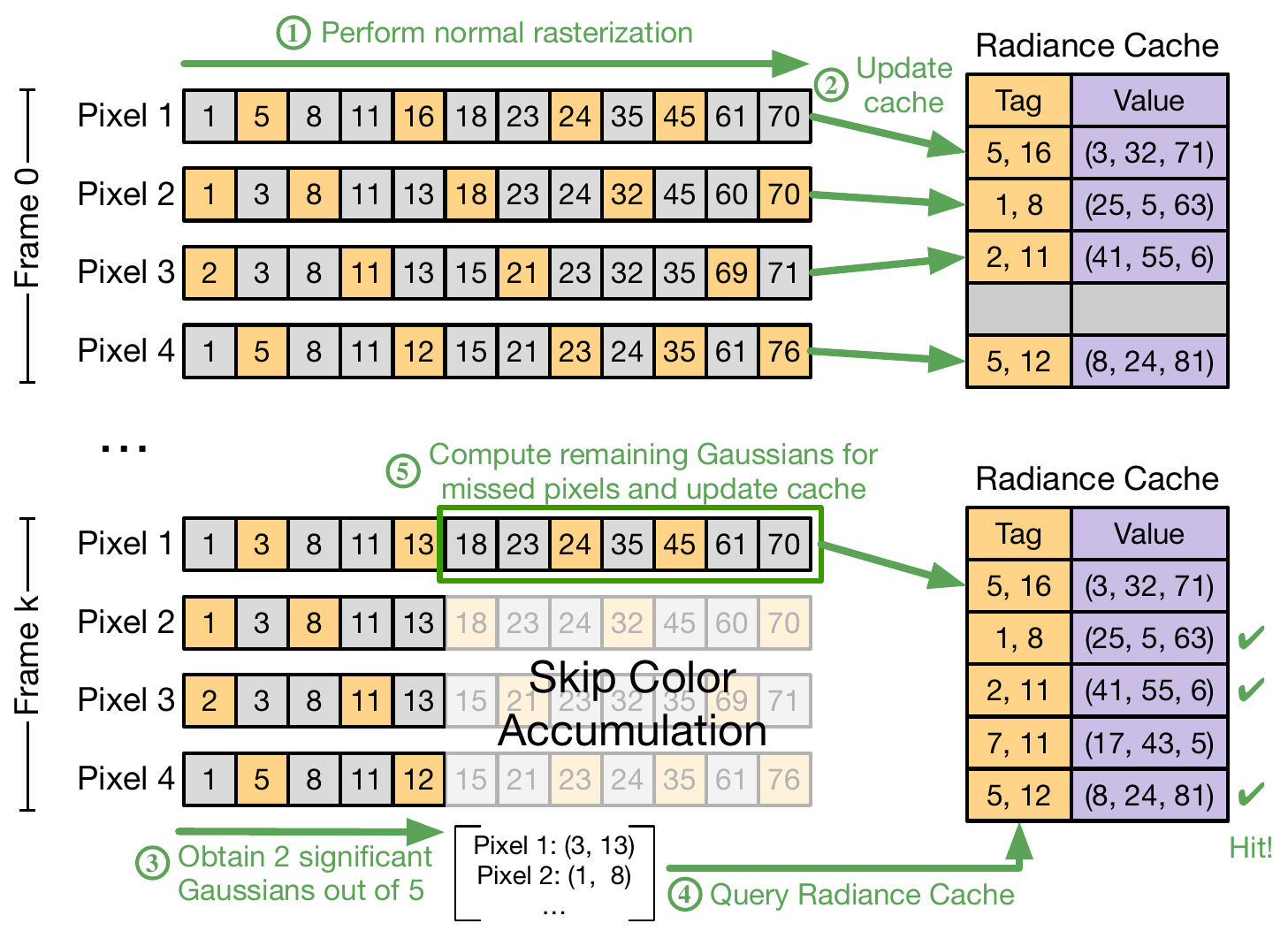}
\caption{
The overall procedure of cached Rasterization step. 
At the initial frame, normal Rasterization is performed, and the radiance cache is populated. 
For each pixel, the cache tag is composed of the first two significant Gaussian IDs (with $\alpha$ > $\frac{1}{255}$ shown in yellow), and the cache value is the pixel value. 
In subsequent frames, each pixel only needs to identify the first two significant Gaussians to query the radiance cache. 
Cache-hit pixels skip the remaining Rasterization.
Cache-missed pixels continue the remaining Rasterization and update the radiance cache accordingly.
}
\label{fig:cached_rasterization}
\end{figure}

\paragraph{Algorithm.} Next, we explain our algorithm, radiance caching, using a toy example illustrated in \Fig{fig:cached_rasterization}:

\circled{white}{1} In the original 3DGS algorithm, each pixel iterates through a set of Gaussians and integrates Gaussians' contributions in order according to \Eqn{eqn:nerf}. 
At the first frame $0$, because the radiance cache is empty, our \rc algorithm performs the exact Rasterization as in the original 3DGS, computing pixel values from scratch.
 
\circled{white}{2} After each pixel is computed, we concatenate the IDs of the first two intersected significant Gaussians (those with $\alpha$ values > $\frac{1}{255}$ highlighted in yellow in \Fig{fig:cached_rasterization}) as the cache tag, and use the computed pixel value serves as the cache value. 
Then, the radiance cache is updated accordingly in \Fig{fig:cached_rasterization}, where the initially empty cache is populated by four pixels. 
We later explain why we select significant Gaussians instead of all Gaussians for caching.

\circled{white}{3} After rendering the first frame and updating the radiance cache, we now can accelerate the rendering of subsequent frames using the radiance cache. 
For instance, when rendering frame $k$, each pixel computes the initial five Gaussians to identify the first two significant Gaussians.

\circled{white}{4} With the significant Gaussian IDs identified, each pixel concatenates these two Gaussian IDs as the cache tag and queries the radiance cache. 
If the tag matches any entry in the cache (e.g., for pixels 2, 3, and 4), those cache-hit pixels can directly use the cached pixel values, and their remaining color integration can be skipped.
In this case, pixels 2, 3, and 4 bypass the color integration of the remaining seven Gaussians.
 
\circled{white}{5} For cache-missed pixels (e.g., pixel 1), they are still required to iterate through the remaining Gaussians to obtain the final pixel values. 
Once the color integration is completed, the radiance cache is then updated using the new values of these cache-missed pixels according to its cache policy.

Overall, our \rc algorithm leverages the inherent color integration process in 3DGS and introduces a caching technique, which identifies similar rays during the initial process of Rasterization and eliminates redundant computations.
Importantly, in our \rc algorithm, only the first frame requires the complete 3DGS rendering, while all subsequent frames benefit from \rc.

\paragraph{Why Significant Gaussians?}
There are two reasons why we choose the first few significant Gaussian IDs as cache tags instead of any arbitrary Gaussian IDs.

\begin{figure}[t]
\centering
\begin{minipage}[t]{0.48\columnwidth}
  \centering
  \includegraphics[width=\columnwidth]{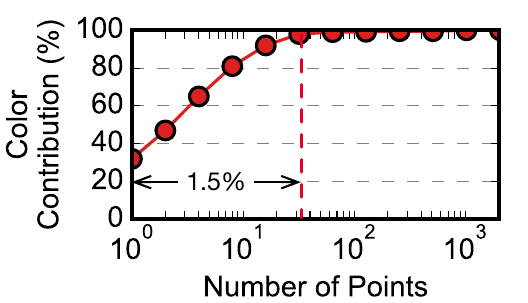}
  \caption{The significance of Gaussian points towards the final radiance. Points are sorted by their contributions.}
  \label{fig:pixel_contribution}
\end{minipage}
\hspace{2pt}
\begin{minipage}[t]{0.48\columnwidth}
  \centering
  \includegraphics[width=\columnwidth]{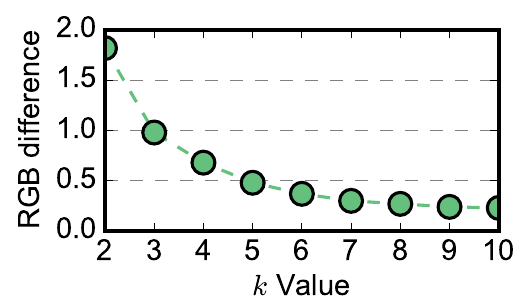}
  \caption{The average color difference between Gaussians that share the same initial significant Gaussians, $k$.}
  \label{fig:path_similarity}
\end{minipage}
\end{figure}

First, significant Gaussians contribute more toward the final pixel values, making them an effective indicator of ray similarity.
As discussed in \Sect{sec:mot:perf}, 
only about 10\% of Gaussians contribute to the final pixel value due to the sparsity of color integration.
We further examine the significance of Gaussians towards the final pixel value in \Fig{fig:pixel_contribution}, where Gaussians are sorted by their contribution. 
The result shows that over 99\% of the pixel value is derived from less than 1.5\% of the Gaussians.
This indicates that the remaining 98.5\% of Gaussians have a negligible impact on rendering quality.

Second, we aim to use the fewest possible Gaussians to index the cache.
Naturally, using initial significant Gaussians as a cache tag allows for a compact cache tag while avoiding the remaining color integration.
We analyzed the average color difference between pixels that share the same initial significant Gaussians. 
In \Fig{fig:path_similarity}, increasing the number of initial significant Gaussians, $k$, reduces the color difference between pixels. 
Specifically, the average color difference remains below 1.0 when $k$ is set to 3, and falls below 0.5 when $k$ increases to 5. 
Both values are negligible compared to the maximum value of 255.
This demonstrates that the initial significant Gaussians are a reliable metric for ray similarity.


\paragraph{General Applicability.} Although we propose radiance caching to accelerate 3DGS, this technique is \textit{not} limited to 3DGS alone.
Radiance caching leverages the fundamental concept in neural rendering: representing scenes with trainable primitives (e.g., Gaussians in 3DGS or voxels in NeRF) and using rays to intersect with these primitives and integrate colors.
Thus, even as neural rendering evolves toward new primitives, our technique remains beneficial, i.e., using the first $k$ number of ``new primitives'' for caching, as long as this core rendering principle holds.

\subsection{Putting it Together}
\label{sec:algo:sys}

\begin{figure}[t]
\centering
\includegraphics[width=\columnwidth]{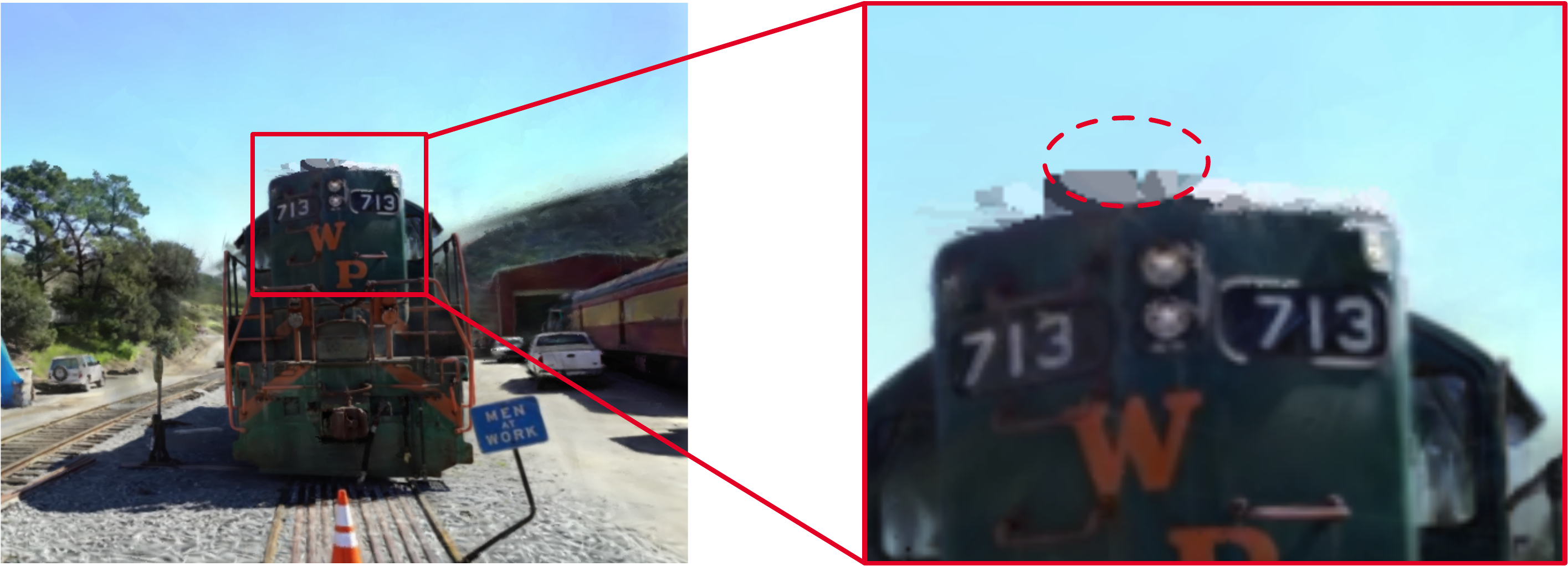}
\caption{An example of the imprecise rendering without cache-aware fine-tuning. The artifact on the top of the train is due to large Gaussians.}
\label{fig:artifact}
\end{figure}

\begin{figure*}[t]
\centering
\includegraphics[width=\textwidth]{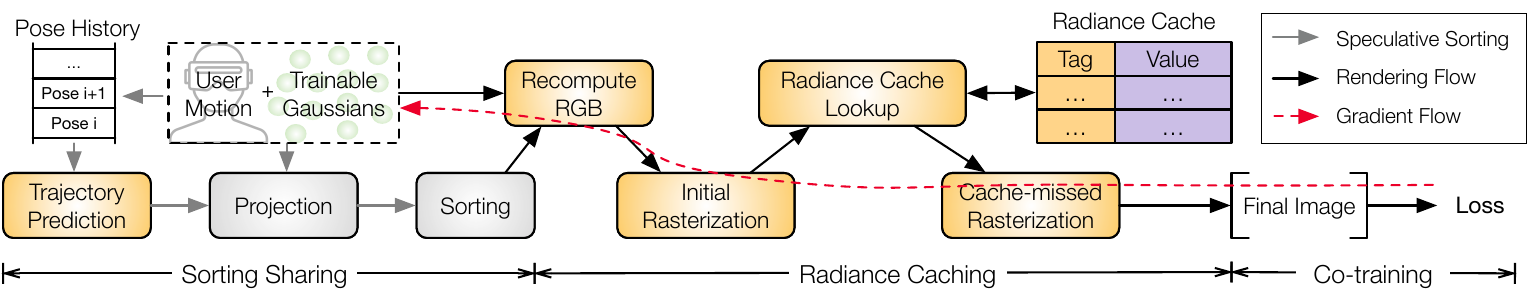}
\caption{The overall \sys system. Our \algo algorithm predicts the future pose and proactively computes sorting which is later shared across multiple frames. Meanwhile, our radiance caching accelerates the Rasterization step by exploiting similarities across rays. The framework is end-to-end trainable, with sorting and cache lookup untouched by gradient descent.}
\label{fig:sys}
\end{figure*}

\paragraph{Overview.} 
\Fig{fig:sys} illustrates the overall workflow of \sys. 
Our system begins by taking the user’s pose history to predict future camera poses. 
\sys then performs Projection and Sorting at the recently predicted pose and saves the sorting results for subsequent frames. 
When \sys receives a user pose update, it directly uses the recent sorting result and recomputes the RGB value of each Gaussian based on the current camera pose before Rasterization. 
During Rasterization, each ray identifies the first $k$ significant Gaussians and queries the radiance cache using these Gaussian IDs. 
Cache-hit pixels would terminate Rasterization early and use the cached pixel values directly, whereas cache-missed pixels continue the full Rasterization process and update the radiance cache accordingly. 
The final image is constructed by combining both cache-hit and cache-missed pixels.

\paragraph{End-to-End Fine-Tuning.}
One key assumption of radiance caching is that the first few significant Gaussian points are sufficiently small and can be a good approximator of the ray similarity.
In cases where the Gaussian points are too large, this assumption may lead to imprecise rendering as shown in \Fig{fig:artifact}. 
There are noticeable artifacts due to the large Gaussians on the top of the train. 
We propose a technique called cache-aware fine-tuning which is designed to enhance the accuracy of radiance caching in these cases.
In particular, we introduce a scale-constrained loss function during training, which constrains the scale of each Gaussian:
\begin{align}
\label{eqn:loss}
   L_{total} = L_{orig} + \alpha * L_{scale} (S, \theta)
\end{align}
\noindent where $L_{orig}$ is the original 3DGS loss and $L_{scale}$ is scale-constrained loss. $S$ is the geometric mean of three Gaussian point scale parameters.
$L_{scale}$ penalties the geometric mean of any Gaussian point greater than $\theta$.
\Sect{sec:eval:acc} shows the effectiveness of scale-constrained loss.
It is worth mentioning that both sorting and cache lookup do not participate in gradient descent as the red dashed line shows.
Therefore, the model is end-to-end differentiable even though sorting and cache lookup are not.

%% file: arch.tex
\section{\core}
\label{sec:arch}

\begin{figure}[t]
\centering
\includegraphics[width=\columnwidth]{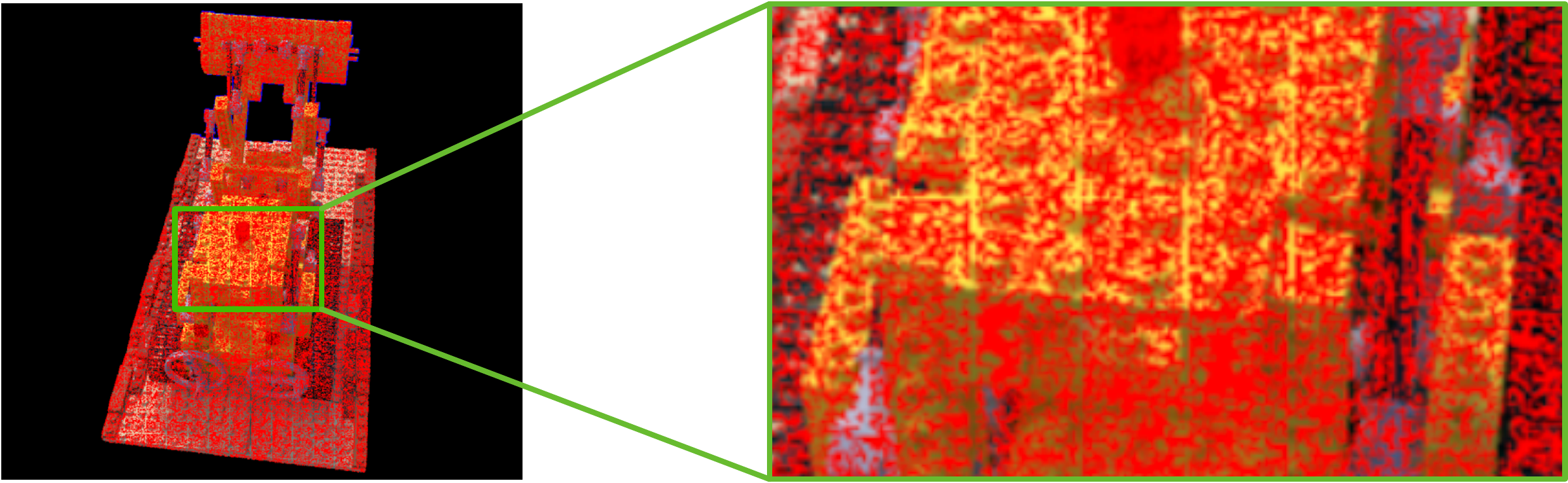}
\caption{An example of the sparse distribution of cache-hit pixels. The red dots highlight the cache-hit pixels which are uniformly distributed across the entire image.}
\label{fig:cache_hit_exp}
\end{figure}

\paragraph{Why Accelerator?} 
While radiance caching (\Sect{sec:algo:rc}) reduces computations in Rasterization, it cannot eliminate Rasterization. 
Recall, \Sect{sec:mot:perf} shows that GPUs are fundamentally ill-suited for Rasterization due to warp divergence. 
Radiance-cached Rasterization exacerbates the divergence, transforming a dense pixel rendering into a sparse one in \Fig{fig:cache_hit_exp}.
The red dots highlight the cache-hit pixels, which are uniformly distributed across the scene.
This further increases the percentage of masked threads after cache lookup.

In the current GPU’s execution model, assigning one thread per pixel requires completed threads to wait until all threads in the warp finish, leading to severe GPU under-utilization.
In addition to warp divergence, implementing radiance caching directly on GPU also introduces additional caching overhead and lock contentions.
Our results in \Sect{sec:eval:perf} will show that the GPU implementation of \rc slows down Rasterization rather than speeding it up.
Thus, we propose dedicated hardware to address the inefficiencies of sparse Rasterization, which is exacerbated by \rc, on GPUs.

\begin{figure*}[t]
\centering
\includegraphics[width=\textwidth]{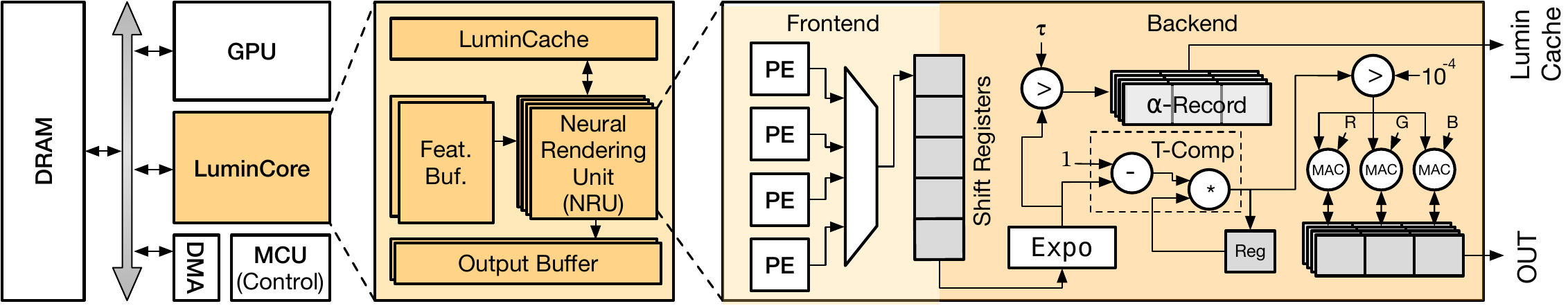}
\caption{The overall SoC architecture design. We integrate our \core into a mobile SoC architecture. The uncolored is the baseline architecture. \core consists of four main components: a \cache, a set of Neural Rendering Units (NRUs), a double-buffered Feature, and an Output Buffer. NRUs execute Rasterization and are designed in a dual-segment fashion to support both dense and sparse populated Rasterization.}
\label{fig:arch}
\end{figure*}

\paragraph{Overview.} Our \core architecture is integrated with a mobile SoC to support the cached Rasterization. 
\Fig{fig:arch} describes our SoC architecture, comprising two primary components: a mobile GPU and our \core. 
During rendering, Rasterization is delegated to \core, while the mobile GPU executes Projection and Sorting. 
\Fig{fig:arch} highlights \core (in color) from the baseline hardware. 

\core includes double-buffered Feature and Output Buffers, which store Gaussian features and the output pixel values, respectively. 
Meanwhile, \core is coupled with a local cache, \cache, tailored for radiance caching.
\cache is shared across multiple Neural Rendering Units (NRUs) to accelerate Rasterization.

\begin{figure}[t]
\centering
\includegraphics[width=\columnwidth]{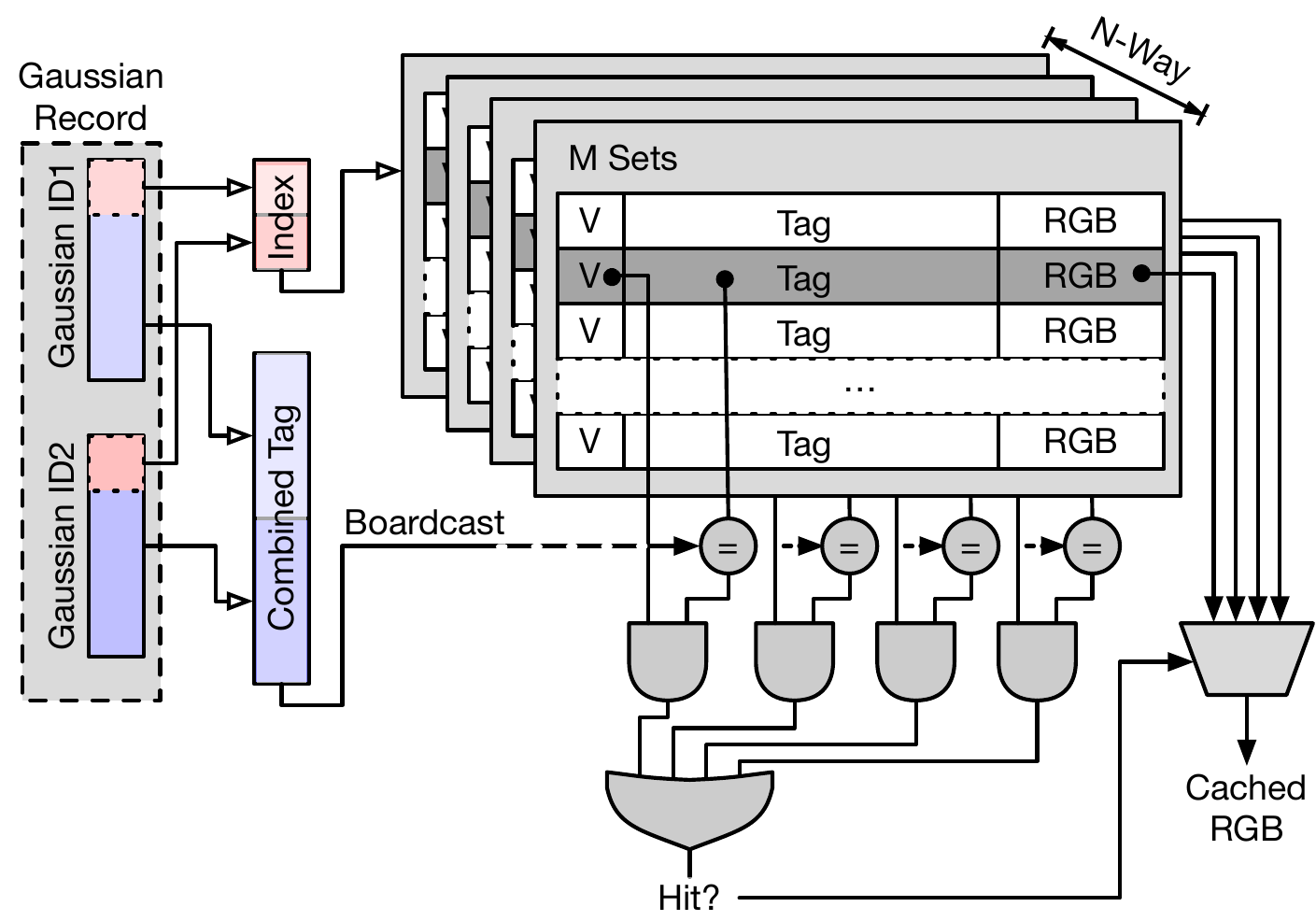}
\caption{Hardware support for radiance caching. The lower bits of the first $k$ significant Gaussians are combined to form an index for accessing the cache. Meanwhile, their higher bits are concatenated to serve as a tag and compared against the tags stored within the cache. If a cache hit happens, the corresponding cached pixel value is returned.
Here, $k$ is set to 2 for illustration purposes.
}
\label{fig:cache}
\end{figure}

\paragraph{\cache.} 
\Fig{fig:cache} illustrates our cache architecture designed to accelerate the cache lookup in \Sect{sec:algo:rc}.
Our cache design resembles a classic $N$-way set associative cache, with $N$ set to 4 in this example.
However, \cache differs in both the indexing method and the cache tag and value design. 

Recall, \rc algorithm requires identifying the first $k$ significant Gaussians ($k$ is set to 2 in this toy example) and using these Gaussian IDs to find the ray with the same initial Gaussian IDs in the cache.
The index of each query is formed by concatenating the lower bits of Gaussian IDs as shown in \Fig{fig:cache}.
The higher bits of all Gaussian IDs are concatenated to form a combined tag to validate the cache hit.
If there is a cache hit, the radiance cache directly returns the cached RGB value to that pixel. 
In the case of a cache miss, the cache-missed pixel continues the remaining Rasterization. 
Once completed, the pixel updates the radiance cache according to a widely-used pseudo least-recently-used (LRU) cache policy~\cite{kkedzierski2010adapting}.

Note that, we design the entire cache to be shared across $2 \times 2$ image tiles.
Rendering the next batch of $2 \times 2$ tiles requires first saving the current cache data to memory, flushing the entire cache, and loading data related to the new batch from memory.
When we render the same $2 \times 2$ tiles in the next frame, we first reload the corresponding cache data from the memory and then perform rasterization. 
In addition, our cache is designed to be double-buffered to hide the latency of data loading.

\paragraph{Neural Rendering Unit.}
GPU is inefficient for radiance-cached Rasterization primarily due to the warp divergence, as only a small number of Gaussians have low transparency and thus contribute to the color integration process.
Prior 3DGS accelerators~\cite{lee2024gscore, feng2024potamoi} inherit the same inefficiency.
To address this issue, our idea is to decouple transparency computation and color integration into two separate algorithmic stages, each of which is scheduled to fully utilize the corresponding hardware structures.


In particular, our NRU is divided into a frontend for calculating Gaussian transparency and a backend for performing color integration.
The backend is designed to be shared across multiple processing elements (PEs) in the frontend so that we can ensure the full utilization of the backend despite the sparsity inherent in color integration.

The frontend is responsible for the lightweight computation that would apply to all Gaussians.
It consists of a set of PEs, where each PE computes the transparency of Gaussians for each pixel. 
These PEs check whether the transparency is significant enough (i.e., $\alpha$ > 1/255) to affect the pixel’s final value, ensuring that only significant Gaussians would be inserted into a FIFO implemented by shift registers for color integration.

\begin{figure}[t]
\centering
\includegraphics[width=\columnwidth]{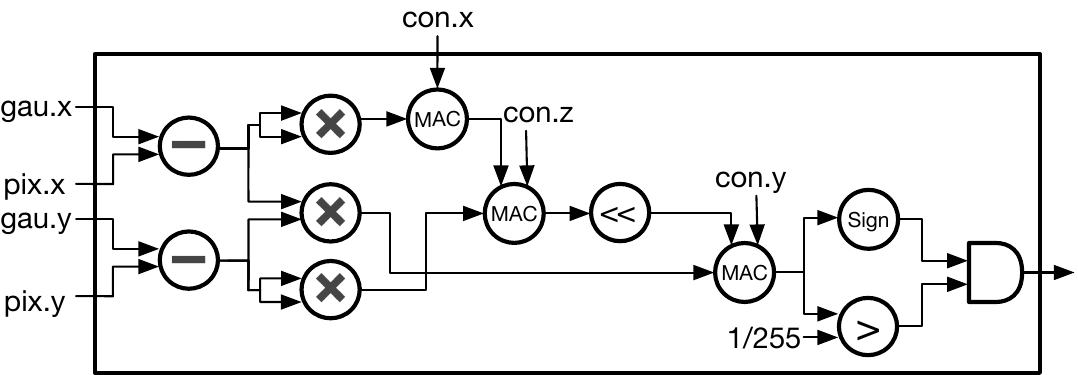}
\caption{The detailed architecture of a three-stage PE. Each PE calculates the exponential decay and assesses the significance of each Gaussian before sending it to the backend.}
\label{fig:pe}
\end{figure}

The detailed design of PE is described in \Fig{fig:pe}. 
Overall, each PE is implemented as a three-stage pipeline.
Each PE is implemented with three multipliers and three multiply-and-add (MAC) units to compute the $\alpha$ values of Gaussians. 
Additionally, each PE uses a comparator and a sign checker to assess whether the Gaussian's $\alpha$ is significant enough to grant subsequent processing.

The backend, which is shared among multiple PEs, is for compute-intensive but sparse color integration that is applied only to significant Gaussians, including dedicated components for computing exponent, etc. 
Each time, the backend takes one Gaussian from the FIFO and performs color integration for its corresponding pixels.
The backend also contains a set of register files ($\alpha$-records) to cache significant Gaussian IDs for different pixels which will be used for radiance cache lookup. 
This frontend-backend design improves PE utilization by leveraging the sparsity of significant Gaussians.

\paragraph{Sparsity-Aware Remapping.} 
Recall that radiance caching saves computations of cache-hit pixels, while the cache-miss pixels still require full rasterization.
{\Fig{fig:cache_hit_exp}} shows that these cache-miss pixels are sparsely distributed across the entire image.
Retaining the same pixel-to-PE mapping, where each PE is responsible for a single pixel, would lead to PE under-utilization, because the cache-hit PEs would remain idle after cache lookup.
To improve the PE utilization, we allow the NRUs to be reconfigured so that all the PEs within a single NRU can collaborate on rendering a \textit{single} pixel.

For all the PEs to collaboratively render the same pixel in this mode, the PEs read different Gaussians (that belong to a pixel/tile) from a sorted Gaussian list in order, instead of reading the same Gaussian (to process different pixels in parallel).
Then, the PEs would write the per-Gaussian intermediate results (e.g., transparency, Gaussian ID) into the shift registers in order if the Gaussian transparencies are so high that require further processing by the backend.
The backend executes in the same way as in the normal mode.
By doing so, we avoid PE under-utilization once some pixels are terminated early by radiance caching.


\paragraph{SoC Integration.}
{$\core$} is a standalone SoC IP block and can be integrated well with different GPU architectures. 
The data communication between {$\core$} and a GPU is via standard SoC-level interconnect (e.g., AXI).
In our case, there is no direct interaction between {$\core$} and the GPU.
{$\core$} only read data from DRAM through DMA.
Thus, our design is agnostic to and can accommodate different GPU architectures.

\paragraph{Broad Applicability.}
Note that, {$\core$} accelerates the general field of point-based neural rendering, which has wide applicability not only in rendering but also in areas like SLAM\mbox{~\cite{matsuki2024gaussian, yan2024gs}} and beyond.
With the recent rise of neural rendering, rendering primitives are rapidly evolving from voxels~{\cite{mildenhall2021nerf}} to 3D Gaussians~{\cite{kerbl20233d}} and 2D Gaussians~{\cite{huang20242d}}.
Despite this rapid development, the overall rendering process, i.e., color accumulation, stays the same. 
Our {$\core$} holds the potential to be used beyond 3DGS, as we extend to new rendering primitives.
In addition, {$\core$} essentially performs parallel and sparse accumulation, a computing paradigm that is also prevalent in other domains such as sparse linear algebra~{\cite{duff2002overview}}, graph neural networks~{\cite{wu2020comprehensive}}, and more.
By making {$\core$} programmable, we can support other domains as well.

%% file: exp.tex
\section{Experimental Setup}
\label{sec:exp}

\paragraph{Hardware Setup.} The \core has $8\times8$ NRUs clocked at 1~GHz, each consists of four three-stage PEs.
The feature buffer, shared across all NRUs, is double-buffered with a total size of 176~KB.  
The output buffer is also double-buffered with a size of 6~KB. Each NRU includes a 160~B shift register array to store temporal $\alpha$ values of significant Gaussians, with 88~B of register files to store $\alpha$ records.

The \cache is shared among the NRUs. 
This cache is designed as a 4-way associative cache, comprising $4\times1024$ entries with a total size of 52~KB. 
The tag of each cache entry is constructed by the IDs of 5 significant Gaussians (using the 3rd to 18th least significant bits of each Gaussian ID, 10 bytes in total), with the corresponding RGB color values as cache values. 
Overall, \cache can cache $64\times64$ pixels to share across $4\times4$ image tiles with a size of $16\times16$.
\cache is also double-buffered to hide the latency of loading cached values from previous renderings.

\paragraph{Experimental Methodology.} The performance of the GPU, including kernel launch times, is directly measured on the mobile Volta GPU in Nvidia’s Xavier SoC. 
Power metrics are also directly obtained using the built-in power measurement features of the device~\cite{xaviersoc}.
The \core’s datapath is developed using an EDA process that includes synthesis, placement, and routing with Synopsys and Cadence tools on TSMC’s 16 nm FinFET technology. 
These results are then scaled to the 12 nm node of Nvidia’s Xavier SoC using DeepScaleTool to be compatible with the Nvidia GPU~\cite{stillmaker2017scaling, sarangi2021deepscaletool}. 
SRAM components are produced using the Arm Artisan memory compiler, with power estimates determined via Synopsys PrimeTimePX with annotated switching activities.

\paragraph{Simulation Methodology.} We simulate the entire system with our cycle-accurate simulator, which is implemented with component-level latencies and power measurements.
The latency and energy of NRU are obtained from the post-synthesis results of its RTL design and scaled down to 12 nm node using DeepScaleTool~{\cite{sarangi2021deepscaletool}} to match the mobile Volta GPU on Nvidia Xavier SoC{~\cite{xaviersoc}}.
The latency of GPU execution is directly measured, including kernel launch time.
The GPU power consumption is obtained via the built-in power measurement on Nvidia Xavier SoC.

The DRAM model in our simulation is based on Micron’s 16 Gb LPDDR3-1600, utilizing four channels according to its datasheet~{\cite{micronlpddr3}}, with energy consumption sourced from Micron System Power Calculators~{\cite{microdrampower}}. 
The energy ratio between a random DRAM access and an SRAM access is about 25:1 aligned with prior work\mbox{\cite{gao2017tetris, Yazdanbakhsh2018GAN}}.

The system energy is the sum of GPU, {$\core$}, and DRAM. 
The overall latency is derived from the combined execution of GPU, NRU, and DRAM.
Note that, the total latency excludes the parallel execution between sorting on GPU and rasterization on NRU.
Due to the double buffering in NRU and {$\cache$}, the overall latency is dominated by the compute latency, not memory.

\paragraph{Area Overhead.} \proj introduces minimal area overhead with the \core design, primarily due to the $8\times8$ NRU array. 
The area overhead, amounting to 1.05 mm$^2$, is negligible when compared to the entire mobile SoC area, which is approximately 350 mm$^2$ for Nvidia’s Xavier SoC~\cite{xaviersochotchips}. 

\paragraph{Datasets.} To evaluate the efficiency and robustness of \algo and \rc, we evaluate both synthetic and real-world scenes. 

For synthetic scenes, we select four out of eight scenes from Synthetic-NeRF (S-NeRF)~\cite{mildenhall2021nerf}.
We use the raw Blender files to generate videos and simulate a typical VR scenario with the average head rotation of 25 degrees at 90 FPS~\cite{visionprospec, questprospec, hendicott2002head, wang2023effect}.

In addition, we use Tanks\&Temples (T\&T)~\cite{Knapitsch2017}, a real-world dataset, from which we choose four sequences for our experiments. 
We extract a 10-second video clip from the raw video sequence in each sequence and use COLMAP~\cite{schoenberger2016sfm}, a well-known photogrammetry tool, to generate the necessary camera poses for our evaluations. 
Note that, the raw videos in T\&T are captured at 30 FPS, much lower than the 90 FPS typically used in VR scenarios.
The rendering quality of \algo and \rc is assessed using Peak Signal-to-Noise Ratio (PSNR), SSIM, and LPIPS as the standard metrics.
We cannot evaluate our techniques on MipNeRF360 (U360){~\cite{barron2022mipnerf360}} and DeepBlending (DB){~\cite{hedman2018deep}} datasets used in the original 3DGS paper because they contain individual images, \textit{not} continuous video sequences.

\paragraph{Hardware Baselines.}
We compare two hardware baselines in our evaluation. 
One is the mobile Volta GPU in Nvidia’s Xavier SoC. 
The other, \mode{NRU+GPU}, is the SoC architecture in \Fig{fig:arch}, but excluding \cache. 
Both baselines execute the full-fledged 3DGS algorithm. 
\mode{NRU+GPU} executes Projection and Sorting on GPU while executing Rasterization on NRU.

\paragraph{Variants.}
To dissect our contributions in algorithm and architecture, we evaluate five variants of \proj to separate the contributions proposed in our paper:
\begin{itemize}
\item \mode{\algo-GPU}: executes the \algo algorithm on a mobile Volta GPU with radiance caching disabled.
\item \mode{\rc-GPU}: executes the original 3DGS algorithm with \rc mechanism on a mobile Volta GPU.
\item \mode{\algo-Acc}: executes the \algo algorithm on our proposed architecture with radiance caching disabled.
\item \mode{\rc-Acc}: executes the original 3DGS algorithm with \rc mechanism on our proposed architecture.
\item \mode{\proj}: the full-fledged \proj, with both \algo algorithm and \rc mechanism on our architectures.
\end{itemize}

\paragraph{User Study.} We also conduct a user study to assess the subjective rendering quality of our techniques, the procedure is approved by our Internal Review Board (IRB). 
Our user study includes 30 participants (20 males and 10 females; age 18-30).
No participants were aware of the research objectives, experimental hypothesis, or the number of conditions. 
All participants had normal or corrected-to-normal vision.
Each participant reviews 4 traces in our evaluation.
We use the classic Two-Interval Forced Choice (2IFC) procedure~\cite{bogacz2006physics}.
For each trace, we display the renderings of these two methods side-by-side on a 16-inch monitor in a randomized order to each participant, with 30 seconds to rest and answer questions.
We ask two questions to each participant:
\begin{itemize}
    \item Whether they notice any difference between these two.
    \item If they observe any difference, pick one that they prefer.
\end{itemize}
If participants do not notice any difference, they are required to choose the version they prefer.
Each trace is repeated three times, and the trace order is also randomized each time. 

%% file: eval.tex
\section{Evaluation}
\label{sec:eval}

We first demonstrate that \proj achieves comparable rendering quality against the baseline both qualitatively and quantitatively (\Sect{sec:eval:acc}). Next, we show the speedup and energy reduction of \proj against two hardware baselines (\Sect{sec:eval:perf}). We then conduct a sensitivity study on \proj (\Sect{sec:eval:sens}). 
Finally, we show that \proj performs better against GSCore~\cite{lee2024gscore} (\Sect{sec:eval:gscore}).
Finally, we show that \proj achieves better performance compared to the state-of-the-art 3DGS accelerator, GSCore~\cite{lee2024gscore} (\Sect{sec:eval:gscore}).

\subsection{Rendering Quality}
\label{sec:eval:acc}

\begin{figure}[t]
\centering
\subfloat[Question one: any difference between two versions. Numbers show participants who notice no difference.]{
	\label{fig:q1}	
        \includegraphics[width=0.47\columnwidth]{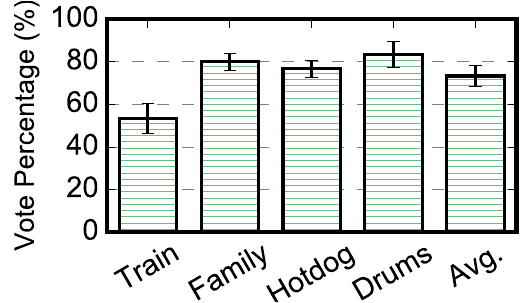} 
}
\hspace{2pt}
\subfloat[Question two: pick one you prefer. The percentages show participants who prefer our method.]{
	\label{fig:q2}
	\includegraphics[width=0.47\columnwidth]{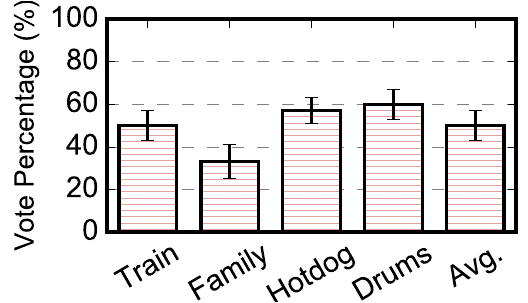} 
 } 
\caption{User study of \proj compared against original 3DGS. Among 27\% of participants who notice differences, we achieve a 50\%-50\% tie.}
\vspace{-5pt}
\label{fig:user_study}
\end{figure}

\paragraph{Subjective Evaluation.} 
\Fig{fig:q1} shows the percentage of votes that do not notice any difference from our user study.
On average, over 73\% of users do not notice any difference between our method and the baseline 3DGS.
\Fig{fig:q2} shows the percentage of votes that notice any difference and prefer our method.
Among the remaining 27\% of participants who noticed differences, 50\% preferred the renderings by \proj.
This shows that our rendering method has an equal visual quality compared to the baseline 3DGS.
We also set up a website to show our rendering results: \href{https://lumina-project.netlify.app/}{link}.

\begin{figure}[t]
\centering
\subfloat[Synthetic scenes on PSNR. Higher is better.]{
	\label{fig:synthetic_acc}	
        \includegraphics[width=0.98\columnwidth]{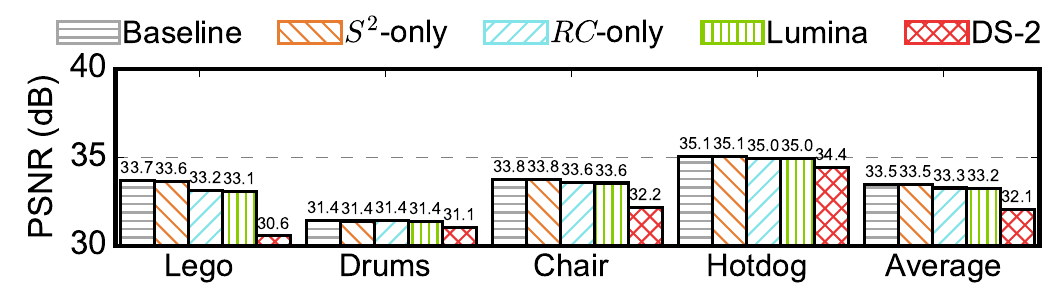}}
\\
\subfloat[Real scenes on PSNR. Higher is better.]{
	\label{fig:real_world_acc}
	\includegraphics[width=0.98\columnwidth]{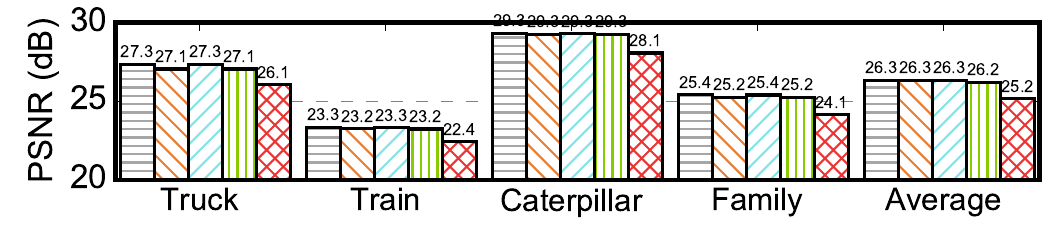}} 
\\
\subfloat[Synthetic scenes on SSIM. Higher is better.]{
	\label{fig:synthetic_ssim}	
        \includegraphics[width=0.98\columnwidth]{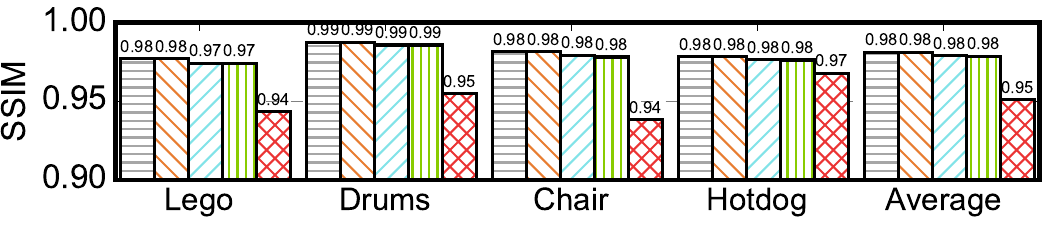}}
\\
\subfloat[Real scenes on SSIM. Higher is better.]{
	\label{fig:real_world_ssim}
	\includegraphics[width=0.98\columnwidth]{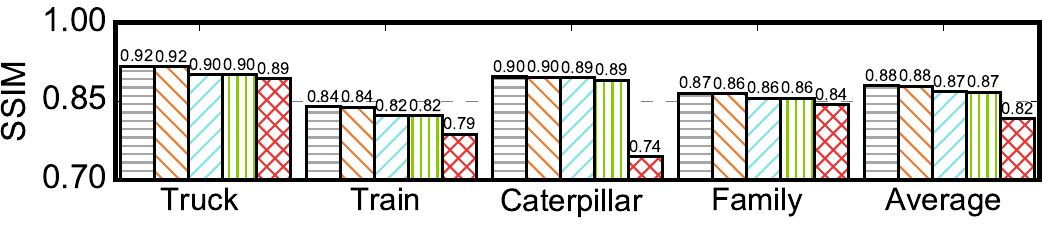}} 
\\
\subfloat[Synthetic scenes on LPIPS. Lower is better.]{
	\label{fig:synthetic_lpips}	
        \includegraphics[width=0.98\columnwidth]{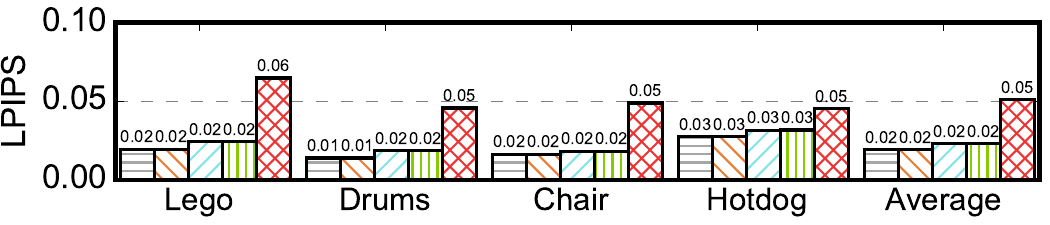}}
\\
\subfloat[Real scenes on LPIPS. Lower is better.]{
	\label{fig:real_world_lpips}
	\includegraphics[width=0.98\columnwidth]{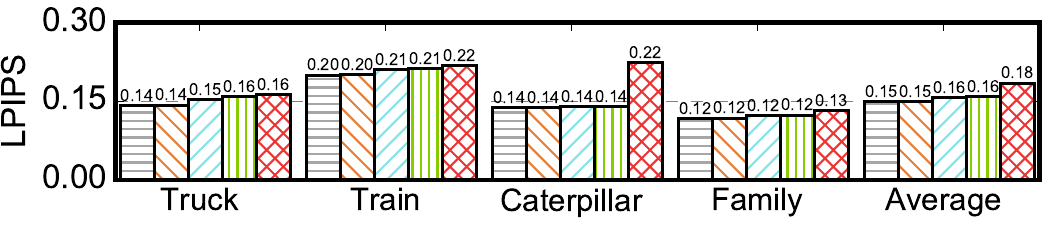}} 
\caption{Image quality comparison. Both \mode{\algo-only} and \mode{\proj} configure the expanded margin (i.e., the number of pixels that the sorting viewport expands) of 4 and the skipping window (i.e., the number of frames share a single sorting result) of 6. \mode{DS-2} first renders a 2$\times$ downsampled frame and then upsamples it to the original resolution by 2.
}
\label{fig:acc}
\end{figure}

\begin{figure}[t]
\centering
\subfloat[Quality evaluation.]{
	\label{fig:scale_loss_acc}	
        \includegraphics[width=0.49\columnwidth]{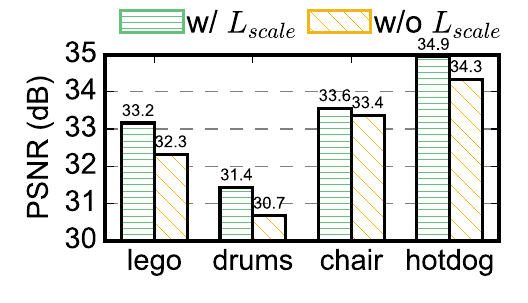} 
}
\subfloat[Cache hit rate.]{
	\label{fig:scale_loss_cache}
	\includegraphics[width=0.49\columnwidth]{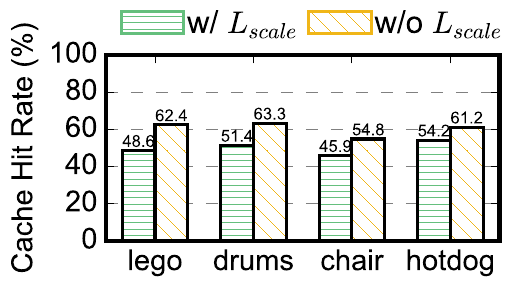} 
 } 
\caption{Rendering quality and cache hit rate of \mode{$\cRC$-only} with and without scale-constrained loss, $L_{scale}$, in \Sect{sec:algo:sys}.}
\label{fig:scale_loss}
\end{figure}

\paragraph{Quantitative Evaluation.} \Fig{fig:acc} compares the rendering quality of various methods on both synthetic and real-world scenes. 
We use \textit{sharing window} to denote the number of frames that share a single sorting result.
We use \textit{expanded margin} to denote the number of pixels that the sorting viewport expands at each dimension (in \Sect{sec:algo:ss}).
Unless specified otherwise, the sharing window is set to be 6, with an expanded margin of 4.
In addition to the baseline algorithm, we compare against a variant, \mode{DS-2}, which first renders a 2$\times$ downsampled frame and then upsamples it to the original resolution by 2. \mode{\algo-only} and \mode{\rc-only} denote only apply \algo algorithm and \rc mechanism, respectively.

\paragraph{Synthetic Scenes.} \Fig{fig:synthetic_acc} shows the rendering quality on synthetic scenes.
Overall, \mode{\algo-only} demonstrates robust performance, matching the baseline in accuracy while only requiring sorting once every 6 frames.
Both \mode{$\cRC$-only} and \mode{\proj} can maintain a similar accuracy across all scenes, with minimal quality losses of $0.2$~dB and $0.3$~dB, respectively.
In Comparison, \mode{DS-2} generally shows a substantial drop in PSNR, with an average $1.4$~dB accuracy drop.

\paragraph{Real Scenes.} 
\Fig{fig:real_world_acc} shows the quality evaluation on real scenes. 
Unlike the synthetic scenes, the real-world scenes have a lower frame rate (30~FPS) than a typical VR scenario (90~FPS), resulting in larger inter-frame movements.
Consequently, \mode{\algo-only} shows a slight decrease in rendering quality by 0.1~dB compared with the baseline. 
Interestingly, \mode{$\cRC$-only} achieves the same accuracy as the baseline, showing the resilience of radiance caching even at lower frame rates.
Overall, \mode{\proj} manages to align closely with the baseline.
In contrast, \mode{DS-2} is 1.0~dB lower than the baseline, vastly underperforming compared with other variants.

{\Fig{fig:synthetic_ssim}} to {\Fig{fig:real_world_lpips}} show the rendering quality of synthetic and real-world scenes on the other two metrics, SSIM and LPIPS. 
The overall trend still holds.
{\mode{\proj}} can achieve a similar rendering quality compared to the corresponding baselines.

\paragraph{Cache-Aware Finetuning.} \Fig{fig:scale_loss} compares the effects of incorporating a scale-constrained loss, $L_{scale}$, in \Sect{sec:algo:sys} on both the rendering quality and cache hit rate on \mode{$\cRC$-only}.
Introducing $L_{scale}$ yields a notable improvement in PSNR across various scenes, with an average improvement of 0.6~dB. 
Including $L_{scale}$ marginally decreases the cache hit rate, which might lead to slight performance degradation, as shown in \Fig{fig:scale_loss_cache}. 
The decrease in cache hit rate is attributed to more stringent constraints on Gaussian point scaling imposed by $L_{scale}$, leading to fewer cache hits.
A quantitative study of the relationship between cache hit rate and performance is presented in \Sect{sec:eval:sens}.

\subsection{Performance and Energy}
\label{sec:eval:perf}

\begin{figure}[t]
\centering
\subfloat[Normalized speedup. Higher is better.]{
    \label{fig:speedup}	
    \includegraphics[width=\columnwidth]{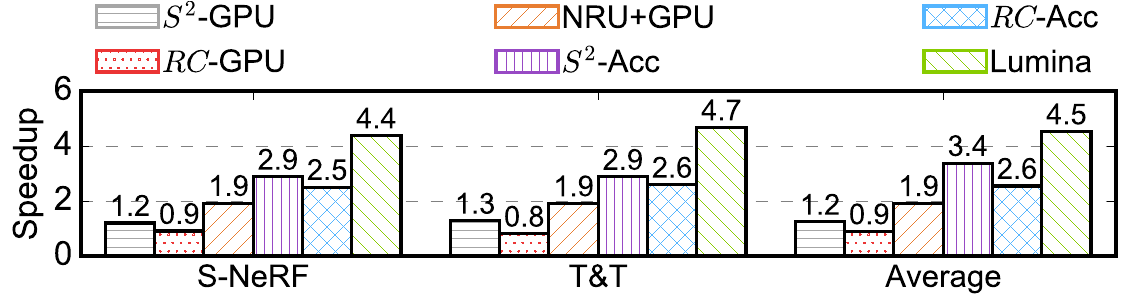}
}
\\
\subfloat[Normalized energy. Lower is better.]{
    \label{fig:energy}
    \includegraphics[width=\columnwidth]{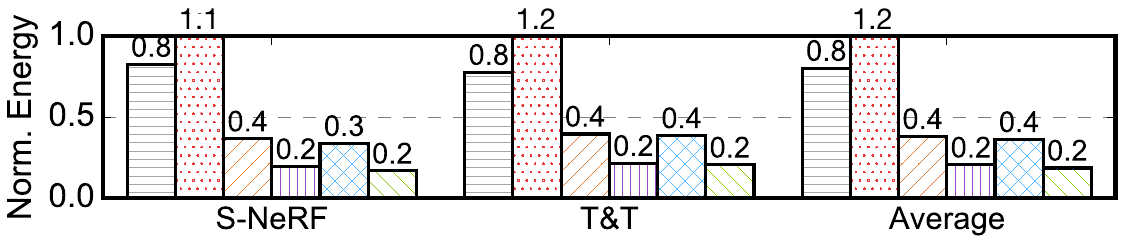}
} 
\caption{Speedup and normalized energy consumptions of different variants over GPU baseline described in \Sect{sec:exp}.}
\label{fig:perf}
\end{figure}

\paragraph{Performance.} 
\Fig{fig:speedup} shows the normalized speedup of different variants compared to the GPU baseline across scenes.
With pure software implementation of \algo and \rc, \mode{\algo-GPU} can achieve 1.2$\times$ speedup by skipping Projection and Sorting.
In comparison, \mode{\rc-GPU} slows down the overall rendering process despite achieving over 50\% cache hit rate.
This slowdown is primarily attributed to GPU warp divergence, as discussed in \Sect{sec:arch}, where the inefficiencies in sparse Rasterization negate the benefits of caching.

With the integration of NRU, \mode{NRU+GPU} yields 1.9$\times$ speedup against the GPU baselines by accelerating the Rasterization stage.
On average, our \core accelerates the Rasterization stage itself by 6.4$\times$.
On top of that, \mode{\algo-Acc} further enhances performance by skipping the execution of the Projection and Sorting stages, boosting the performance to 3.1$\times$.
Since \mode{NRU+GPU} already reducing the execution latency of Rasterization, \mode{\rc-Acc} shows a moderate speedup on top of \mode{NRU+GPU}. 
The overall speedup of \mode{\rc-Acc} ranges from 1.7$\times$ to 2.7$\times$ by leveraging \cache to reduce computational redundancy further.
On the Rasterization stage alone, \mode{\rc-Acc} can achieve 2.5$\times$ speedup.
Together, \mode{\proj} achieves 4.5$\times$ speedup against the GPU baseline across scenes.
On average, {\mode{\proj}} achieves 218.5 FPS and 97.9 FPS on synthetic and real-world scenes, respectively.

\paragraph{Energy Reduction.} The normalized energy consumption of various variants are illustrated in \Fig{fig:energy}. 
Similar to the performance result, \mode{\algo-GPU} achieves 20\% of energy saving while \mode{\rc-GPU} introduces an additional 20\% energy overhead.
By solely leveraging the \core architecture, \mode{NRU+GPU} already demonstrates 62\% of energy reduction, drastically reducing the latency of Rasterization. 
With Rasterization being accelerated, \mode{\algo-Acc} outperforms \mode{$\cRC$-Acc} in terms of energy savings, achieving 79\% and 64\% energy efficiency, respectively. 
Overall, by integrating all techniques, \mode{\proj} achieves energy savings by 81\%, highlighting the synergistic benefits of combining \algo algorithm and \rc mechanism.
Note that, only {\mode{\proj}} achieves real-time (90 FPS) on the real-world dataset. If we set the performance target to be real-time, the energy savings of {\mode{\proj}} would be 93\% and 80\% over the baseline on the synthetic and real-world scenes, respectively.

\begin{figure}[t]
\centering
\subfloat[Quality evaluation.]{
	\label{fig:s2_acc}	
        \includegraphics[width=0.48\columnwidth]{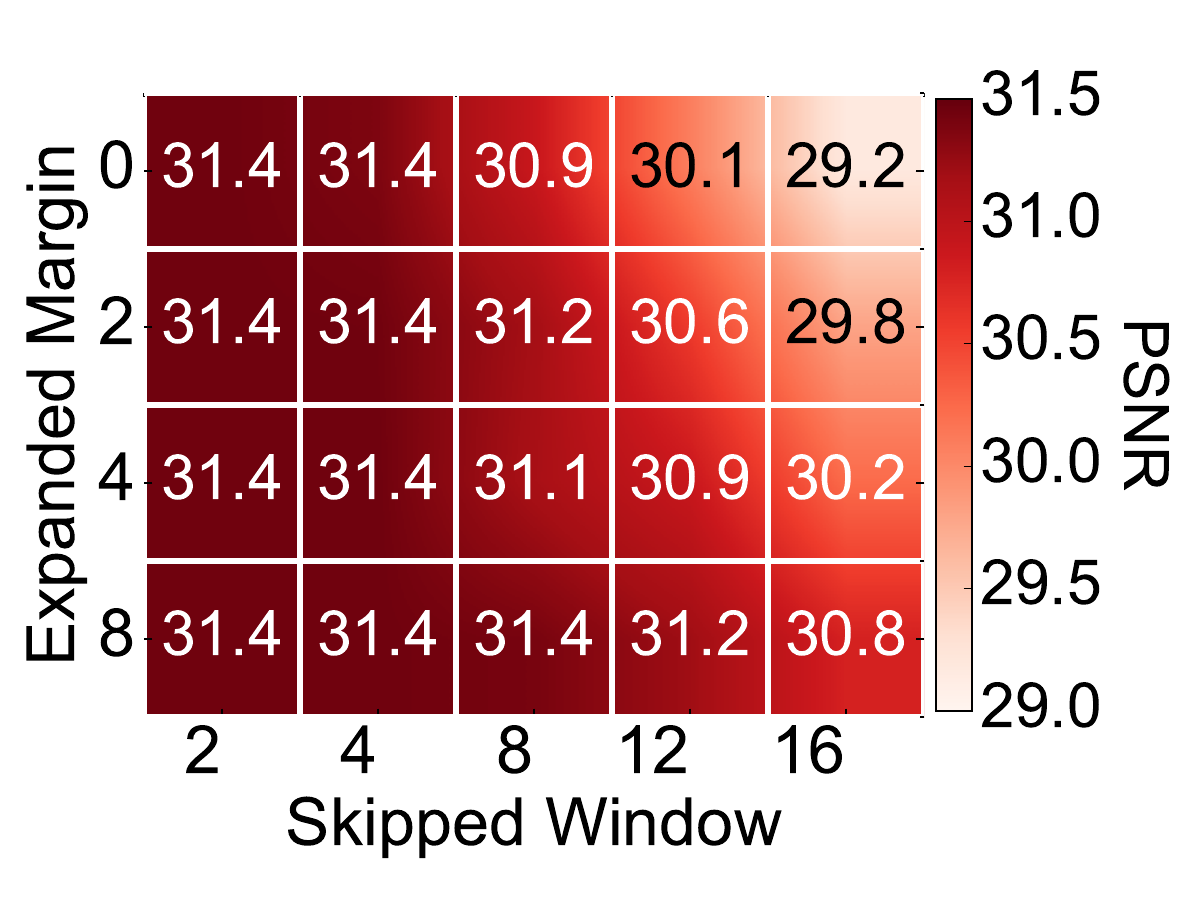} 
}
\subfloat[Normalized speedup.]{
	\label{fig:s2_speedup}
	\includegraphics[width=0.48\columnwidth]{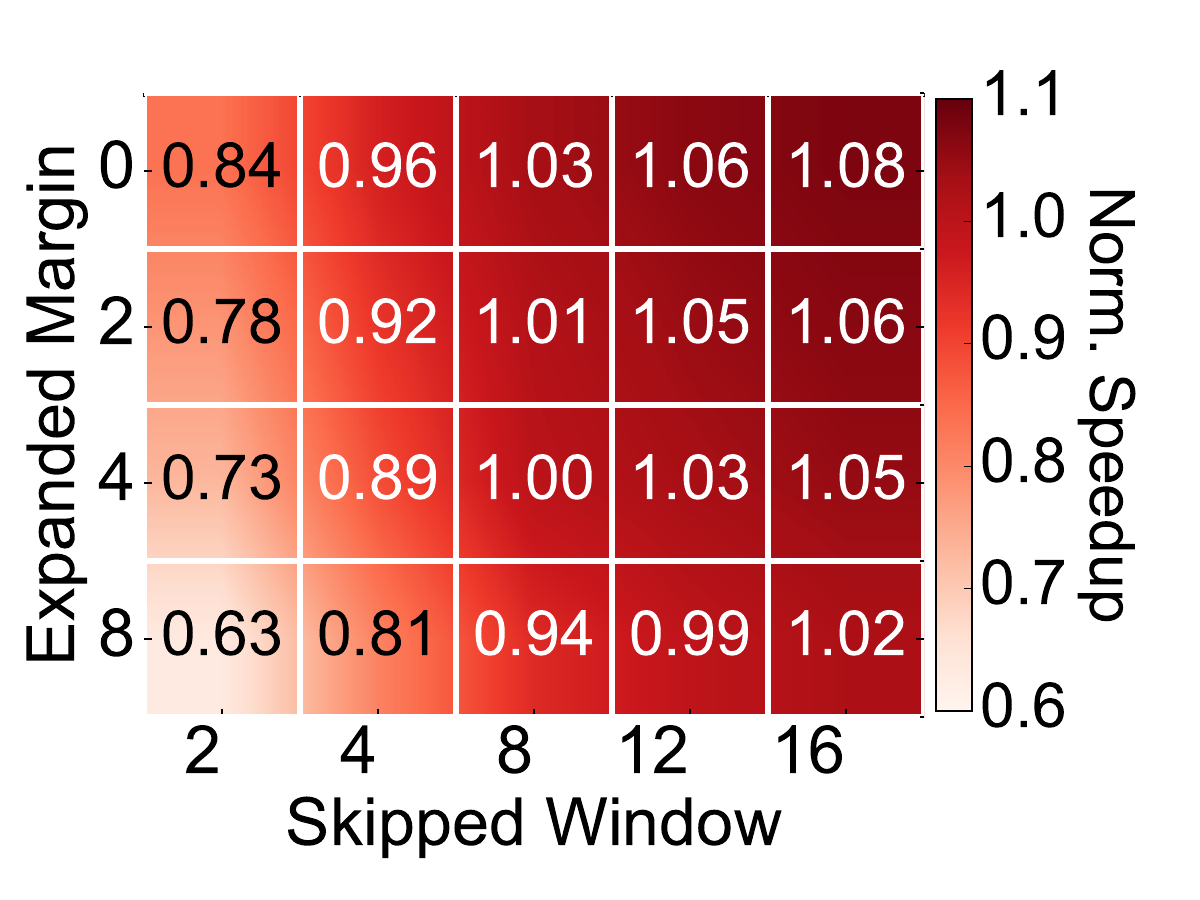} 
 } 
\caption{Sensitivity of rendering quality and performance to the expanded margin and skipped window of the \algo algorithm on the `Drums' scene in SyntheticNeRF dataset. The speedup is normalized to \mode{\algo-only} with an expanded margin of 4 and a skipped frame of 6.
}
\label{fig:s2_sens}
\end{figure}

\subsection{Sensitivity Study}
\label{sec:eval:sens}

\Fig{fig:s2_sens} illustrates the sensitivity of rendering quality and normalized speedup to two algorithmic configurations, expanded viewport and skipped window, on the \textit{drums} scene from the SyntheticNeRF dataset. 
We use \textit{expanded margin} to denote the number of pixels that the sorting viewport expands at each dimension (in \Sect{sec:algo:ss}).
The speedup is normalized to \mode{\algo-only} with an expanded margin of 4 and a skipped window of 6.

\paragraph{Expanded Viewport.} \Fig{fig:s2_acc} shows that, as the expanded margin increases, there is an improvement in rendering quality. 
For instance, with a skipped window of 8, the rendering quality starts at 30.9 dB (with an expanded margin of 2) and increases to 31.4 dB as the expanded margin expands to 8. 
However, increasing the expanded margin results in a speedup decrease, as shown in \Fig{fig:s2_speedup}.
Initially, at an expanded margin of 2, the speedup ranges from 0.8 to 1.1$\times$ compared to the baseline.
As the expanded margin increases to 8, the speedup diminishes to $0.6$ - $1.0\times$, indicating a trade-off between efficiency and quality.

\paragraph{Skipped Window.} Meanwhile, as the number of skipped frames increases, the rendering quality tends to decrease. For example, at the expanded margin of 4, PSNR drops from 31.4 dB to 30.2 dB when the number of skipped frames increases from 2 to 16. Conversely, the speedup increases with more skipped frames. For instance, at the expanded margin of 2, the speedup enhances from 0.8 to 1.1$\times$ with more frames being skipped.

\begin{figure}[t]
\centering
\begin{minipage}[t]{0.48\columnwidth}
  \centering
  \includegraphics[width=\columnwidth]{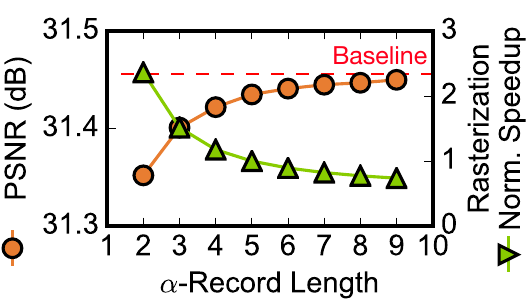}
  \caption{The sensitivity of rendering quality and normalized speedup to the number of significant Gaussians.}
  \label{fig:rc_sens}
\end{minipage}
\hspace{2pt}
\begin{minipage}[t]{0.48\columnwidth}
  \centering
  \includegraphics[width=\columnwidth]{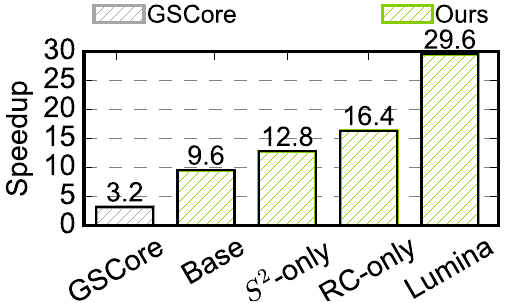}
  \caption{Comparison of speedup between \proj and GSCore. Values are normalized to the GPU baseline.}
  \label{fig:compare_gscore}
\end{minipage}
\end{figure}

\paragraph{$\alpha$-Record Length.} 
\Fig{fig:rc_sens} exploits the sensitivity of rendering quality and normalized speedup to variations in $\alpha$-record length, ranging from 1 to 10. 
Recall, $\alpha$-record stores the IDs of significant Gaussians.
By default, $\alpha$-record length is 5.
Here, we normalize the speedup to the $\alpha$-record length of 5 and only show the normalized speedup in terms of the Rasterization stage. 

As the $\alpha$-record length increases, the rendering quality, represented by the orange circles, gradually increases but eventually plateaus, aligning with the baseline (red dashed line) at 31.4 dB. 
Meanwhile, the normalized speedup of Rasterization, shown by green triangles, shows a gradual decrease from 2.3$\times$ to 0.7$\times$. 
This decrease in speedup is primarily due to a reduction in cache hit rate, which falls from 82\% to 31\%, and an increase in the computational workload before the cache lookup as the $\alpha$-record length extends.

\subsection{Comparison with GSCore}
\label{sec:eval:gscore}

\Fig{fig:compare_gscore} compares the speedup of \proj against the state-of-the-art accelerator, GSCore~\cite{lee2024gscore}. 
For a fair comparison, we incorporate the dedicated accelerator units: Culling\& Conversion Unit (CCU) and Gaussian Sorting Unit (GSU) from GSCore.
In our baseline hardware, projection and sorting are executed on CCU and GSU, respectively, rather than being performed on the GPU.

In \Fig{fig:compare_gscore}, the results are obtained from both synthetic and real-world datasets, with all values normalized to the GPU baseline.
Notably, \proj significantly outperforms GSCore across all variants.
Even our baseline hardware (9.6$\times$) can achieve better performance against GSCore (3.2$\times$), thanks to the frontend-backend design of our \core.
Most of the Gaussian points can avoid unnecessary color integration.
On top of that, \mode{\algo-only} and \mode{$\cRC$-only} can reach 12.8$\times$ and 16.4$\times$.
Here, \mode{$\cRC$-only} achieves a higher speedup compared to \mode{\algo-only} as Rasterization now becomes the dominant stage.
With all the techniques combined, \mode{\proj} can achieve 29.6$\times$ speedup against the GPU baseline.

%% file: related.tex
\section{Related Work}
\label{sec:related}

\paragraph{Neural Rendering Acceleration.} 
In recent years, neural rendering has gained significant attention, leading to the development of numerous accelerators specifically designed for neural rendering algorithms~\cite{lee2023neurex, rao2022icarus, li2023instant, mubarik2023hardware, li2022rt, fu2023gen, feng2024cicero, feng2024potamoi, lin2025metasapiens}. 
Despite that, most prior designs focus exclusively on NeRF and not on 3DGS, which remains less explored. 
Some proposed software-based acceleration techniques, such as pruning~\cite{fang2024mini, fan2023lightgaussian} and quantization~\cite{lee2023compact}, seek to enhance the 3DGS performance but still fall short of achieving real-time performance on mobile devices. 
By far, GScore~\cite{lee2024gscore} remains the only accelerator dedicated to 3DGS. 

This paper introduces techniques that are generally applicable across different 3DGSs and are orthogonal to the approaches used in GScore. 
Our caching can potentially extend its applications beyond 3DGS to enhance NeRFs.

\paragraph{Radiance Caching.}
Prior studies~\cite{krivanek2005radiance, scherzer2012pre, vardis2014real, muller2021real} have explored radiance caching to accelerate ray tracing, but their approach differs fundamentally from ours.
Their techniques primarily focus on caching radiance samples from ray-object intersections.
Their primary focus is to reduce the caching overhead~\cite{vardis2014real, scherzer2012pre} and enable the lightweight computation to radiance interpolation from cached samples using spherical harmonics~\cite{krivanek2005radiance} or neural network~\cite{muller2021real}.

Applying conventional radiance caching to 3DGS would introduce significant storage overhead with no computational savings, as 3DGS does not require multi-bounce irradiance collection~\cite{jones2016parallel}. 
Our radiance caching leverages the 3DGS characteristics, where a single ray intersects multiple Gaussians with no bounce. 
This makes \proj a natural fit to accelerate 3DGS.

\paragraph{Temporal Correlation.}
Prior studies leverage the temporal correlations by classic image warping techniques~\cite{chen2023view, chen1995quicktime, chaurasia2020passthrough+}.
Recent hardware-algorithm co-design generalize warping to use temporal correlations across frames for reducing computation in real-time vision~\cite{buckler2018eva2, zhu2018euphrates, feng2019asv, ying2022exploiting, song2020vr, xiao2020neural, feng2023fast, zhao2020deja, zhao2021holoar, feng2022real, lee2025vr}. 
However, these techniques typically exploit task-specific characteristics and cannot be directly applicable to neural rendering. 
The most recent work, Cicero~\cite{feng2024cicero}, requires known object meshes and applies warping techniques to NeRFs but at a cost of rendering quality. 
In contrast, our techniques, \algo and \rc, boost performance with a marginal quality loss.

%% file: discussion.tex
\section{Discussion and Limitations}
\label{sec:disc}

As with all prior work\mbox{~\cite{buckler2018eva2, zhu2018euphrates, feng2019asv, ying2022exploiting, song2020vr}} that leverages the temporal correlations, a pathological case with rapid head rotations would be detrimental to the performance of {\algo}.
To avoid catastrophic cases, we can simply disable {\algo} by detecting the rapid rotation data from IMU. 
Nevertheless, under common scenarios, our results show that {\algo} is effective in real-world cases ({\Fig{fig:acc}}).

GPUs have a tradition of incorporating custom hardware to support new rendering paradigms such as ray tracing.
We set out to understand whether custom hardware can enable real-time 3DGS, an emerging rendering paradigm.
Starting from a clean architectural slate allows us to fully exploit the algorithmic characteristics of 3DGS without being constrained by the current GPU design.
Although it is important to investigate improvements to the GPU architecture for 3DGS and revisit classical techniques that tackle issues like warp divergence \mbox{~\cite{meng2010dynamic, fung2007dynamic, narasiman2011improving}}, the value of work lies in demonstrating how custom hardware for 3DGS might look like \textit{if} it is warranted.
While the jury is still out, we argue that a dedicated accelerator could have its value in freeing up the GPU resources for other rendering/computation workloads, as future mobile SoCs will undoubtedly execute multiple workloads that exercise the GPUs.

%% file: conclusion.tex
\section{Conclusions}
\label{sec:conc}

Neural rendering has revolutionized the landscape of VR/AR and photo-realistic rendering lately, offering unprecedented realism.
3DGS emerges as a promising alternative to the conventional rasterization pipeline.
It is important to rethink and develop brand-new architectures tailored to this future rendering technology.

This paper exploits a key insight, Gaussian-ray intersection in 3DGS, to design a caching mechanism applicable to 3DGS pipelines. 
With our framework, \sys, we achieve a 4.5$\times$ speedup with minimal yet principle hardware augmentations. 
By integrating our design into a dedicated neural rendering accelerator, \proj can further boost the performance up to 30$\times$.

%% file: paper.bbl
\begin{thebibliography}{10}

\bibitem{visionprospec}
{Apple Vision Pro screen refresh rate is up to 100Hz}.

\bibitem{questprospec}
{Meta Quest Pro specs}.

\bibitem{micronlpddr3}
{Micron 178-Ball, Single-Channel Mobile LPDDR3 SDRAM Features}.

\bibitem{microdrampower}
{Micron System Power Calculators}.

\bibitem{xaviersoc}
Nvidia reveals xavier soc details.

\bibitem{xaviersochotchips}
{NVIDIA’s Xavier System-on-Chip, HotChips 30}.

\bibitem{xr2release}
{Qualcomm Powers Next-Gen Spatial Computing With XR2 Gen 2 And AR1 Gen 1 Platforms}.

\bibitem{snapdragonxr2}
{Qualcomm QCS8550/QCM8550 Processors}.

\bibitem{barron2021mip}
Jonathan~T Barron, Ben Mildenhall, Matthew Tancik, Peter Hedman, Ricardo Martin-Brualla, and Pratul~P Srinivasan.
\newblock Mip-nerf: A multiscale representation for anti-aliasing neural radiance fields.
\newblock In {\em Proceedings of the IEEE/CVF International Conference on Computer Vision}, pages 5855--5864, 2021.

\bibitem{barron2022mipnerf360}
Jonathan~T. Barron, Ben Mildenhall, Dor Verbin, Pratul~P. Srinivasan, and Peter Hedman.
\newblock Mip-nerf 360: Unbounded anti-aliased neural radiance fields.
\newblock {\em CVPR}, 2022.

\bibitem{bogacz2006physics}
Rafal Bogacz, Eric Brown, Jeff Moehlis, Philip Holmes, and Jonathan~D Cohen.
\newblock The physics of optimal decision making: a formal analysis of models of performance in two-alternative forced-choice tasks.
\newblock {\em Psychological review}, 113(4):700, 2006.

\bibitem{buckler2018eva2}
Mark Buckler, Philip Bedoukian, Suren Jayasuriya, and Adrian Sampson.
\newblock Eva$^2$: Exploiting temporal redundancy in live computer vision.
\newblock In {\em 2018 ACM/IEEE 45th Annual International Symposium on Computer Architecture (ISCA)}, pages 533--546. IEEE, 2018.

\bibitem{chaurasia2020passthrough+}
Gaurav Chaurasia, Arthur Nieuwoudt, Alexandru-Eugen Ichim, Richard Szeliski, and Alexander Sorkine-Hornung.
\newblock Passthrough+ real-time stereoscopic view synthesis for mobile mixed reality.
\newblock {\em Proceedings of the ACM on Computer Graphics and Interactive Techniques}, 3(1):1--17, 2020.

\bibitem{chen2024survey}
Guikun Chen and Wenguan Wang.
\newblock A survey on 3d gaussian splatting.
\newblock {\em arXiv preprint arXiv:2401.03890}, 2024.

\bibitem{chen2022geometry}
Mingfei Chen, Jianfeng Zhang, Xiangyu Xu, Lijuan Liu, Yujun Cai, Jiashi Feng, and Shuicheng Yan.
\newblock Geometry-guided progressive nerf for generalizable and efficient neural human rendering.
\newblock In {\em European Conference on Computer Vision}, pages 222--239. Springer, 2022.

\bibitem{chen1995quicktime}
Shenchang~Eric Chen.
\newblock Quicktime vr: An image-based approach to virtual environment navigation.
\newblock In {\em Proceedings of the 22nd annual conference on Computer graphics and interactive techniques}, pages 29--38, 1995.

\bibitem{chen2023view}
Shenchang~Eric Chen and Lance Williams.
\newblock View interpolation for image synthesis.
\newblock In {\em Seminal Graphics Papers: Pushing the Boundaries, Volume 2}, pages 423--432. 2023.

\bibitem{chen2023mobilenerf}
Zhiqin Chen, Thomas Funkhouser, Peter Hedman, and Andrea Tagliasacchi.
\newblock Mobilenerf: Exploiting the polygon rasterization pipeline for efficient neural field rendering on mobile architectures.
\newblock In {\em Proceedings of the IEEE/CVF Conference on Computer Vision and Pattern Recognition}, pages 16569--16578, 2023.

\bibitem{duff2002overview}
Iain~S Duff, Michael~A Heroux, and Roldan Pozo.
\newblock An overview of the sparse basic linear algebra subprograms: The new standard from the blas technical forum.
\newblock {\em ACM Transactions on Mathematical Software (TOMS)}, 28(2):239--267, 2002.

\bibitem{fan2023lightgaussian}
Zhiwen Fan, Kevin Wang, Kairun Wen, Zehao Zhu, Dejia Xu, and Zhangyang Wang.
\newblock Lightgaussian: Unbounded 3d gaussian compression with 15x reduction and 200+ fps.
\newblock {\em arXiv preprint arXiv:2311.17245}, 2023.

\bibitem{fang2024mini}
Guangchi Fang and Bing Wang.
\newblock Mini-splatting: Representing scenes with a constrained number of gaussians.
\newblock {\em arXiv preprint arXiv:2403.14166}, 2024.

\bibitem{feng2022real}
Yu~Feng, Nathan Goulding-Hotta, Asif Khan, Hans Reyserhove, and Yuhao Zhu.
\newblock Real-time gaze tracking with event-driven eye segmentation.
\newblock In {\em 2022 IEEE Conference on Virtual Reality and 3D User Interfaces (VR)}, pages 399--408. IEEE, 2022.

\bibitem{feng2023fast}
Yu~Feng, Patrick Hansen, Paul~N Whatmough, Guoyu Lu, and Yuhao Zhu.
\newblock Fast and accurate: Video enhancement using sparse depth.
\newblock In {\em Proceedings of the IEEE/CVF Winter Conference on Applications of Computer Vision}, pages 4492--4500, 2023.

\bibitem{feng2024potamoi}
Yu~Feng, Weikai Lin, Zihan Liu, Jingwen Leng, Minyi Guo, Han Zhao, Xiaofeng Hou, Jieru Zhao, and Yuhao Zhu.
\newblock Potamoi: Accelerating neural rendering via a unified streaming architecture.
\newblock {\em ACM Transactions on Architecture and Code Optimization}, 2024.

\bibitem{feng2024cicero}
Yu~Feng, Zihan Liu, Jingwen Leng, Minyi Guo, and Yuhao Zhu.
\newblock Cicero: Addressing algorithmic and architectural bottlenecks in neural rendering by radiance warping and memory optimizations.
\newblock {\em arXiv preprint arXiv:2404.11852}, 2024.

\bibitem{feng2019asv}
Yu~Feng, Paul Whatmough, and Yuhao Zhu.
\newblock Asv: Accelerated stereo vision system.
\newblock In {\em Proceedings of the 52nd Annual IEEE/ACM International Symposium on Microarchitecture}, pages 643--656, 2019.

\bibitem{fu2023gen}
Yonggan Fu, Zhifan Ye, Jiayi Yuan, Shunyao Zhang, Sixu Li, Haoran You, and Yingyan Lin.
\newblock Gen-nerf: Efficient and generalizable neural radiance fields via algorithm-hardware co-design.
\newblock In {\em Proceedings of the 50th Annual International Symposium on Computer Architecture}, pages 1--12, 2023.

\bibitem{fung2007dynamic}
Wilson~WL Fung, Ivan Sham, George Yuan, and Tor~M Aamodt.
\newblock Dynamic warp formation and scheduling for efficient gpu control flow.
\newblock In {\em 40th Annual IEEE/ACM International Symposium on Microarchitecture (MICRO 2007)}, pages 407--420. IEEE, 2007.

\bibitem{gao2022nerf}
Kyle Gao, Yina Gao, Hongjie He, Dening Lu, Linlin Xu, and Jonathan Li.
\newblock Nerf: Neural radiance field in 3d vision, a comprehensive review.
\newblock {\em arXiv preprint arXiv:2210.00379}, 2022.

\bibitem{gao2017tetris}
Mingyu Gao, Jing Pu, Xuan Yang, Mark Horowitz, and Christos Kozyrakis.
\newblock Tetris: Scalable and efficient neural network acceleration with 3d memory.
\newblock In {\em Proceedings of the 22nd ACM International Conference on Architectural Support for Programming Languages and Operating Systems}, 2017.

\bibitem{han2020megatrack}
Shangchen Han, Beibei Liu, Randi Cabezas, Christopher~D Twigg, Peizhao Zhang, Jeff Petkau, Tsz-Ho Yu, Chun-Jung Tai, Muzaffer Akbay, Zheng Wang, et~al.
\newblock Megatrack: monochrome egocentric articulated hand-tracking for virtual reality.
\newblock {\em ACM Transactions on Graphics (ToG)}, 39(4):87--1, 2020.

\bibitem{hedman2018deep}
Peter Hedman, Julien Philip, True Price, Jan-Michael Frahm, George Drettakis, and Gabriel Brostow.
\newblock Deep blending for free-viewpoint image-based rendering.
\newblock {\em ACM Transactions on Graphics (ToG)}, 37(6):1--15, 2018.

\bibitem{hedman2021baking}
Peter Hedman, Pratul~P Srinivasan, Ben Mildenhall, Jonathan~T Barron, and Paul Debevec.
\newblock Baking neural radiance fields for real-time view synthesis.
\newblock In {\em Proceedings of the IEEE/CVF International Conference on Computer Vision}, pages 5875--5884, 2021.

\bibitem{hendicott2002head}
PL~Hendicott, B~Brown, KL~Schmid, and S~Fisher.
\newblock Head movement amplitude and velocity during a common visual task.
\newblock {\em Investigative Ophthalmology \& Visual Science}, 43(13):4668--4668, 2002.

\bibitem{hu2022efficientnerf}
Tao Hu, Shu Liu, Yilun Chen, Tiancheng Shen, and Jiaya Jia.
\newblock Efficientnerf efficient neural radiance fields.
\newblock In {\em Proceedings of the IEEE/CVF Conference on Computer Vision and Pattern Recognition}, pages 12902--12911, 2022.

\bibitem{huang20242d}
Binbin Huang, Zehao Yu, Anpei Chen, Andreas Geiger, and Shenghua Gao.
\newblock 2d gaussian splatting for geometrically accurate radiance fields.
\newblock In {\em ACM SIGGRAPH 2024 conference papers}, pages 1--11, 2024.

\bibitem{jiang2022neuman}
Wei Jiang, Kwang~Moo Yi, Golnoosh Samei, Oncel Tuzel, and Anurag Ranjan.
\newblock Neuman: Neural human radiance field from a single video.
\newblock In {\em European Conference on Computer Vision}, pages 402--418. Springer, 2022.

\bibitem{jones2016parallel}
Nathaniel~L Jones and Christoph~F Reinhart.
\newblock Parallel multiple-bounce irradiance caching.
\newblock In {\em Computer Graphics Forum}, volume~35, pages 57--66. Wiley Online Library, 2016.

\bibitem{kkedzierski2010adapting}
Kamil K{\k{e}}dzierski, Miquel Moreto, Francisco~J Cazorla, and Mateo Valero.
\newblock Adapting cache partitioning algorithms to pseudo-lru replacement policies.
\newblock In {\em 2010 IEEE International Symposium on Parallel \& Distributed Processing (IPDPS)}, pages 1--12. IEEE, 2010.

\bibitem{kerbl20233d}
Bernhard Kerbl, Georgios Kopanas, Thomas Leimk{\"u}hler, and George Drettakis.
\newblock 3d gaussian splatting for real-time radiance field rendering.
\newblock {\em ACM Transactions on Graphics}, 42(4):1--14, 2023.

\bibitem{kerbl2024hierarchical}
Bernhard Kerbl, Andreas Meuleman, Georgios Kopanas, Michael Wimmer, Alexandre Lanvin, and George Drettakis.
\newblock A hierarchical 3d gaussian representation for real-time rendering of very large datasets.
\newblock {\em ACM Transactions on Graphics (TOG)}, 43(4):1--15, 2024.

\bibitem{Knapitsch2017}
Arno Knapitsch, Jaesik Park, Qian-Yi Zhou, and Vladlen Koltun.
\newblock Tanks and temples: Benchmarking large-scale scene reconstruction.
\newblock {\em ACM Transactions on Graphics}, 36(4), 2017.

\bibitem{krivanek2005radiance}
Jaroslav Kriv{\'a}nek, Pascal Gautron, Sumanta Pattanaik, and Kadi Bouatouch.
\newblock Radiance caching for efficient global illumination computation.
\newblock {\em IEEE Transactions on Visualization and Computer Graphics}, 11(5):550--561, 2005.

\bibitem{lee2023compact}
Joo~Chan Lee, Daniel Rho, Xiangyu Sun, Jong~Hwan Ko, and Eunbyung Park.
\newblock Compact 3d gaussian representation for radiance field.
\newblock {\em arXiv preprint arXiv:2311.13681}, 2023.

\bibitem{lee2023neurex}
Junseo Lee, Kwanseok Choi, Jungi Lee, Seokwon Lee, Joonho Whangbo, and Jaewoong Sim.
\newblock Neurex: A case for neural rendering acceleration.
\newblock In {\em Proceedings of the 50th Annual International Symposium on Computer Architecture}, pages 1--13, 2023.

\bibitem{lee2025vr}
Junseo Lee, Jaisung Kim, Junyong Park, and Jaewoong Sim.
\newblock Vr-pipe: Streamlining hardware graphics pipeline for volume rendering.
\newblock {\em arXiv preprint arXiv:2502.17078}, 2025.

\bibitem{lee2024gscore}
Junseo Lee, Seokwon Lee, Jungi Lee, Junyong Park, and Jaewoong Sim.
\newblock Gscore: Efficient radiance field rendering via architectural support for 3d gaussian splatting.
\newblock In {\em Proceedings of the 29th ACM International Conference on Architectural Support for Programming Languages and Operating Systems, Volume 3}, pages 497--511, 2024.

\bibitem{li2022rt}
Chaojian Li, Sixu Li, Yang Zhao, Wenbo Zhu, and Yingyan Lin.
\newblock Rt-nerf: Real-time on-device neural radiance fields towards immersive ar/vr rendering.
\newblock In {\em Proceedings of the 41st IEEE/ACM International Conference on Computer-Aided Design}, pages 1--9, 2022.

\bibitem{li2023instant}
Sixu Li, Chaojian Li, Wenbo Zhu, Boyang Yu, Yang Zhao, Cheng Wan, Haoran You, Huihong Shi, and Yingyan Lin.
\newblock Instant-3d: Instant neural radiance field training towards on-device ar/vr 3d reconstruction.
\newblock In {\em Proceedings of the 50th Annual International Symposium on Computer Architecture}, pages 1--13, 2023.

\bibitem{liang2023envidr}
Ruofan Liang, Huiting Chen, Chunlin Li, Fan Chen, Selvakumar Panneer, and Nandita Vijaykumar.
\newblock Envidr: Implicit differentiable renderer with neural environment lighting.
\newblock {\em arXiv preprint arXiv:2303.13022}, 2023.

\bibitem{lin2025metasapiens}
Weikai Lin, Yu~Feng, and Yuhao Zhu.
\newblock Metasapiens: Real-time neural rendering with efficiency-aware pruning and accelerated foveated rendering.
\newblock In {\em Proceedings of the 30th ACM International Conference on Architectural Support for Programming Languages and Operating Systems, Volume 1}, pages 669--682, 2025.

\bibitem{lindell2022bacon}
David~B Lindell, Dave Van~Veen, Jeong~Joon Park, and Gordon Wetzstein.
\newblock Bacon: Band-limited coordinate networks for multiscale scene representation.
\newblock In {\em Proceedings of the IEEE/CVF conference on computer vision and pattern recognition}, pages 16252--16262, 2022.

\bibitem{liu2024citygaussian}
Yang Liu, He~Guan, Chuanchen Luo, Lue Fan, Junran Peng, and Zhaoxiang Zhang.
\newblock Citygaussian: Real-time high-quality large-scale scene rendering with gaussians.
\newblock {\em arXiv preprint arXiv:2404.01133}, 2024.

\bibitem{matsuki2024gaussian}
Hidenobu Matsuki, Riku Murai, Paul~HJ Kelly, and Andrew~J Davison.
\newblock Gaussian splatting slam.
\newblock In {\em Proceedings of the IEEE/CVF Conference on Computer Vision and Pattern Recognition}, pages 18039--18048, 2024.

\bibitem{meng2010dynamic}
Jiayuan Meng, David Tarjan, and Kevin Skadron.
\newblock Dynamic warp subdivision for integrated branch and memory divergence tolerance.
\newblock In {\em Proceedings of the 37th annual international symposium on Computer architecture}, pages 235--246, 2010.

\bibitem{mildenhall2021nerf}
Ben Mildenhall, Pratul~P Srinivasan, Matthew Tancik, Jonathan~T Barron, Ravi Ramamoorthi, and Ren Ng.
\newblock Nerf: Representing scenes as neural radiance fields for view synthesis.
\newblock {\em Communications of the ACM}, 65(1):99--106, 2021.

\bibitem{mubarik2023hardware}
Muhammad~Husnain Mubarik, Ramakrishna Kanungo, Tobias Zirr, and Rakesh Kumar.
\newblock Hardware acceleration of neural graphics.
\newblock In {\em Proceedings of the 50th Annual International Symposium on Computer Architecture}, pages 1--12, 2023.

\bibitem{muller2021real}
Thomas M{\"u}ller, Fabrice Rousselle, Jan Nov{\'a}k, and Alexander Keller.
\newblock Real-time neural radiance caching for path tracing.
\newblock {\em arXiv preprint arXiv:2106.12372}, 2021.

\bibitem{narasiman2011improving}
Veynu Narasiman, Michael Shebanow, Chang~Joo Lee, Rustam Miftakhutdinov, Onur Mutlu, and Yale~N Patt.
\newblock Improving gpu performance via large warps and two-level warp scheduling.
\newblock In {\em Proceedings of the 44th Annual IEEE/ACM International Symposium on Microarchitecture}, pages 308--317, 2011.

\bibitem{rao2022icarus}
Chaolin Rao, Huangjie Yu, Haochuan Wan, Jindong Zhou, Yueyang Zheng, Minye Wu, Yu~Ma, Anpei Chen, Binzhe Yuan, Pingqiang Zhou, et~al.
\newblock Icarus: A specialized architecture for neural radiance fields rendering.
\newblock {\em ACM Transactions on Graphics (TOG)}, 41(6):1--14, 2022.

\bibitem{rojas2023re}
Sara Rojas, Jesus Zarzar, Juan~C P{\'e}rez, Artsiom Sanakoyeu, Ali Thabet, Albert Pumarola, and Bernard Ghanem.
\newblock Re-rend: Real-time rendering of nerfs across devices.
\newblock In {\em Proceedings of the IEEE/CVF International Conference on Computer Vision}, pages 3632--3641, 2023.

\bibitem{sarangi2021deepscaletool}
Satyabrata Sarangi and Bevan Baas.
\newblock Deepscaletool: A tool for the accurate estimation of technology scaling in the deep-submicron era.
\newblock In {\em 2021 IEEE International Symposium on Circuits and Systems (ISCAS)}, pages 1--5. IEEE, 2021.

\bibitem{scherzer2012pre}
Daniel Scherzer, Chuong~H Nguyen, Tobias Ritschel, and Hans-Peter Seidel.
\newblock Pre-convolved radiance caching.
\newblock In {\em Computer Graphics Forum}, volume~31, pages 1391--1397. Wiley Online Library, 2012.

\bibitem{schoenberger2016sfm}
Johannes~Lutz Sch\"{o}nberger and Jan-Michael Frahm.
\newblock Structure-from-motion revisited.
\newblock In {\em Conference on Computer Vision and Pattern Recognition (CVPR)}, 2016.

\bibitem{song2020vr}
Zhuoran Song, Feiyang Wu, Xueyuan Liu, Jing Ke, Naifeng Jing, and Xiaoyao Liang.
\newblock Vr-dann: Real-time video recognition via decoder-assisted neural network acceleration.
\newblock In {\em 2020 53rd Annual IEEE/ACM International Symposium on Microarchitecture (MICRO)}, pages 698--710. IEEE, 2020.

\bibitem{stillmaker2017scaling}
Aaron Stillmaker and Bevan Baas.
\newblock Scaling equations for the accurate prediction of cmos device performance from 180 nm to 7 nm.
\newblock {\em Integration}, 58:74--81, 2017.

\bibitem{tancik2022block}
Matthew Tancik, Vincent Casser, Xinchen Yan, Sabeek Pradhan, Ben Mildenhall, Pratul~P Srinivasan, Jonathan~T Barron, and Henrik Kretzschmar.
\newblock Block-nerf: Scalable large scene neural view synthesis.
\newblock In {\em Proceedings of the IEEE/CVF Conference on Computer Vision and Pattern Recognition}, pages 8248--8258, 2022.

\bibitem{vardis2014real}
K~Vardis, G~Papaioannou, and A~Gkaravelis.
\newblock Real-time radiance caching using chrominance compression.
\newblock {\em Journal of Computer Graphics Techniques Vol}, 3(4), 2014.

\bibitem{wang2023effect}
Jialin Wang, Rongkai Shi, Wenxuan Zheng, Weijie Xie, Dominic Kao, and Hai-Ning Liang.
\newblock Effect of frame rate on user experience, performance, and simulator sickness in virtual reality.
\newblock {\em IEEE Transactions on Visualization and Computer Graphics}, 29(5):2478--2488, 2023.

\bibitem{weng2022humannerf}
Chung-Yi Weng, Brian Curless, Pratul~P Srinivasan, Jonathan~T Barron, and Ira Kemelmacher-Shlizerman.
\newblock Humannerf: Free-viewpoint rendering of moving people from monocular video.
\newblock In {\em Proceedings of the IEEE/CVF conference on computer vision and pattern Recognition}, pages 16210--16220, 2022.

\bibitem{wu2024recent}
Tong Wu, Yu-Jie Yuan, Ling-Xiao Zhang, Jie Yang, Yan-Pei Cao, Ling-Qi Yan, and Lin Gao.
\newblock Recent advances in 3d gaussian splatting.
\newblock {\em arXiv preprint arXiv:2403.11134}, 2024.

\bibitem{wu2020comprehensive}
Zonghan Wu, Shirui Pan, Fengwen Chen, Guodong Long, Chengqi Zhang, and S~Yu Philip.
\newblock A comprehensive survey on graph neural networks.
\newblock {\em IEEE transactions on neural networks and learning systems}, 32(1):4--24, 2020.

\bibitem{xiangli2022bungeenerf}
Yuanbo Xiangli, Linning Xu, Xingang Pan, Nanxuan Zhao, Anyi Rao, Christian Theobalt, Bo~Dai, and Dahua Lin.
\newblock Bungeenerf: Progressive neural radiance field for extreme multi-scale scene rendering.
\newblock In {\em European conference on computer vision}, pages 106--122. Springer, 2022.

\bibitem{xiao2020neural}
Lei Xiao, Salah Nouri, Matt Chapman, Alexander Fix, Douglas Lanman, and Anton Kaplanyan.
\newblock Neural supersampling for real-time rendering.
\newblock {\em ACM Transactions on Graphics (TOG)}, 39(4):142--1, 2020.

\bibitem{yan2024gs}
Chi Yan, Delin Qu, Dan Xu, Bin Zhao, Zhigang Wang, Dong Wang, and Xuelong Li.
\newblock Gs-slam: Dense visual slam with 3d gaussian splatting.
\newblock In {\em Proceedings of the IEEE/CVF Conference on Computer Vision and Pattern Recognition}, pages 19595--19604, 2024.

\bibitem{Yazdanbakhsh2018GAN}
Amir Yazdanbakhsh, Kambiz Samadi, Nam~Sung Kim, and Hadi Esmaeilzadeh.
\newblock Ganax: A unified mimd-simd acceleration for generative adversarial networks.
\newblock 2018.

\bibitem{ye2023intrinsicnerf}
Weicai Ye, Shuo Chen, Chong Bao, Hujun Bao, Marc Pollefeys, Zhaopeng Cui, and Guofeng Zhang.
\newblock Intrinsicnerf: Learning intrinsic neural radiance fields for editable novel view synthesis.
\newblock In {\em Proceedings of the IEEE/CVF International Conference on Computer Vision}, pages 339--351, 2023.

\bibitem{ying2022exploiting}
Ziyu Ying, Shulin Zhao, Haibo Zhang, Cyan~Subhra Mishra, Sandeepa Bhuyan, Mahmut~T Kandemir, Anand Sivasubramaniam, and Chita~R Das.
\newblock Exploiting frame similarity for efficient inference on edge devices.
\newblock In {\em 2022 IEEE 42nd International Conference on Distributed Computing Systems (ICDCS)}, pages 1073--1084. IEEE, 2022.

\bibitem{zhang2022ray}
Jian Zhang, Yuanqing Zhang, Huan Fu, Xiaowei Zhou, Bowen Cai, Jinchi Huang, Rongfei Jia, Binqiang Zhao, and Xing Tang.
\newblock Ray priors through reprojection: Improving neural radiance fields for novel view extrapolation.
\newblock In {\em Proceedings of the IEEE/CVF Conference on Computer Vision and Pattern Recognition}, pages 18376--18386, 2022.

\bibitem{zhao2020deja}
Shulin Zhao, Haibo Zhang, Sandeepa Bhuyan, Cyan~Subhra Mishra, Ziyu Ying, Mahmut~T Kandemir, Anand Sivasubramaniam, and Chita~R Das.
\newblock D{\'e}ja view: Spatio-temporal compute reuse for ‘energy-efficient 360 vr video streaming.
\newblock In {\em 2020 ACM/IEEE 47th Annual International Symposium on Computer Architecture (ISCA)}, pages 241--253. IEEE, 2020.

\bibitem{zhao2021holoar}
Shulin Zhao, Haibo Zhang, Cyan~Subhra Mishra, Sandeepa Bhuyan, Ziyu Ying, Mahmut~Taylan Kandemir, Anand Sivasubramaniam, and Chita Das.
\newblock Holoar: On-the-fly optimization of 3d holographic processing for augmented reality.
\newblock In {\em MICRO-54: 54th Annual IEEE/ACM International Symposium on Microarchitecture}, pages 494--506, 2021.

\bibitem{zhu2018euphrates}
Yuhao Zhu, Anand Samajdar, Matthew Mattina, and Paul Whatmough.
\newblock Euphrates: Algorithm-soc co-design for low-power mobile continuous vision.
\newblock {\em arXiv preprint arXiv:1803.11232}, 2018.

\end{thebibliography}
